\documentclass[english,numberedappendix]{emulateapj}
\usepackage[T1]{fontenc}
\usepackage{babel}
\usepackage{array}
\usepackage{amssymb}
\usepackage{graphicx}
\usepackage[unicode=true]
 {hyperref}
\usepackage{breakurl}

\makeatletter

\providecommand{\tabularnewline}{\\}

\usepackage{amsmath}
\slugcomment{}
\shorttitle{Superbursts with hydrogen and helium-rich atmospheres}
\shortauthors{Keek, Heger, \& In 't Zand}

\makeatother

\begin{document}

\title{Superburst models for neutron stars with hydrogen and helium-rich
atmospheres}

\author{L.~Keek}

\affil{National Superconducting Cyclotron Laboratory, Department of Physics
and Astronomy, and Joint Institute for Nuclear Astrophysics, Michigan
State University, East Lansing, MI 48824, USA}

\email{keek@nscl.msu.edu}

\author{A.~Heger}

\affil{School of Physics and Astronomy, University of Minnesota, 116 Church
Street SE, Minneapolis, MN 55455, USA}

\author{J.\,J.\,M.~in~'t~Zand}

\affil{SRON Netherlands Institute for Space Research, Sorbonnelaan 2, 3584
CA Utrecht, The Netherlands }
\begin{abstract}
Superbursts are rare day-long Type I X-ray bursts due to carbon flashes
on accreting neutron stars in low-mass X-ray binaries. They heat the
neutron star envelope such that the burning of accreted hydrogen and
helium becomes stable, and the common shorter X-ray bursts are quenched.
Short bursts reappear only after the envelope cools down. We study
multi-zone one-dimensional models of the neutron star envelope, in
which we follow carbon burning during the superburst, and we include
hydrogen and helium burning in the atmosphere above. We investigate
both the case of a solar composition and a helium-rich atmosphere.
This allows us to study for the first time a wide variety of thermonuclear
burning behavior as well as the transitions between the different
regimes in a self-consistent manner. For solar composition, burst
quenching ends much sooner than previously expected. This is because
of the complex interplay between the $3\alpha$, hot CNO, and CNO
breakout reactions. Stable burning of hydrogen and helium transitions
via marginally stable burning (mHz quasi-periodic oscillations) to
less energetic bursts with short recurrence times. We find a short-lived
bursting mode where weaker and stronger bursts alternate. Eventually
the bursting behavior changes back to that of the pre-superburst bursts.
Because of the scarcity of observations, this transition has not been
directly detected after a superburst. Using the MINBAR burst catalog
we identify the shortest upper limit on the quenching time for 4U~1636--536,
and derive further constraints on the time scale on which bursts return.

\end{abstract}


\section{Introduction}

\label{sec:Introduction}Superbursts are day-long flares observed
from neutron stars in low-mass X-ray binaries (LMXBs; \citealt{Cornelisse2000,Strohmayer2002}),
that are attributed to the unstable thermonuclear burning of a carbon-rich
layer (\citealt{Cumming2001}; see \citealt{Cooper2009} for a discussion
on alternative types of fuel). The thermonuclear runaway starts at
a column depth of $y=10^{11}\,\mathrm{g\, cm^{-2}}$ to $y=10^{12}\,\mathrm{g\, cm^{-2}}$,
close to the outer crust (\citealt{Cumming2006}). Because of the
high temperature dependence of carbon burning (e.g., \citealt{Caughlan1988}),
the ignition is sensitive to crustal heating, and hence superburst
observations place constraints on the combined crustal heating and
neutrino cooling of the crust and core. For example, the superburst
from the classical transient X-ray source 4U~1608--522 requires the
crust to have heated up faster than predicted by current models (\citealt{Keek2008}).
Because of the long typical recurrence time of approximately a year,
superbursts are rare: so far 22 (candidates) have been detected from
13 sources (e.g., \citealt{Keek2008int..work}; see \citealt{Kuulkers2009ATel};
\citealt{Chenevez2011ATel}; \citealt{Zand2011ATel}; \citealt{Serino2012};
\citealt{Asada2011ATel} for recent discoveries; see also Table~\ref{tab:Observational-limits-on}).

The carbon fuel is expected to be produced by thermonuclear burning
of accreted hydrogen and/or helium higher up in the atmosphere (e.g.,
\citealt{Strohmayer2002}). All known superbursting sources exhibit
unstable hydrogen/helium burning observed as short (up to $\sim100\,\mathrm{s}$
duration) Type I X-ray bursts, as well as stable burning, evidenced
by high values of the $\alpha$ parameter, the ratio of the integrated
persistent emission between two bursts and the burst fluence (\citealt{Zand2003}).
This burning takes place at a column depth of $y\simeq10^{8}\,\mathrm{g\, cm^{-2}}$.
Whereas stable hydrogen/helium burning through the hot CNO cycle and
the $3\alpha$-process produce carbon, it is destroyed in bursts by
CNO-cycle breakout reactions, followed by a series of helium captures
catalyzed by protons --- the \textsl{$\alpha$p}-process --- and proton
captures with subsequent $\beta$ decays --- the \textsl{rp}-process
--- creating heavier elements (\citealt{Schatz2001,2003Schatz}).
Current models produce burst ashes with a carbon mass fraction of
typically $5\%$ (\citealt{Woosley2004}), whereas a slightly higher
mass fraction is found for models that accrete material with a higher
metallicity at a lower rate. Superburst observations, however, indicate
a carbon content that is closer to $20\%$ (\citealt{Cumming2006}).
\citet{Medin2011} recently suggested that chemical separation by
freeze out close to the outer crust may increase the carbon fraction.

If the superburst ignites in a sufficiently thick layer, carbon burning
initially proceeds as a detonation (\citealt{Weinberg2006sb}). This
produces a shock that travels to the surface, pushing the envelope
outward. The hydrogen/helium-rich atmosphere is heated by the shock
(\citealt{Weinberg2007}) and the subsequent fall-back of the outer
layers (\citealt{Keek2011}). This produces a bright precursor burst.
Furthermore, the ignition conditions for a hydrogen/helium flash are
reached. X-ray emission from this flash adds to the precursor burst,
or it may account for the entire precursor emission, in case there
is no strong shock (\citealt{Keek2011}). Precursor bursts have been
identified in all cases when the start of the superburst was observed
and the data were of sufficient quality (e.g., \citealt{Kuulkers2003a}).

After the precursor burst, the envelope is sufficiently hot for all
subsequently accreted hydrogen and helium to burn stably at the same
rate at which it is accreted: the bursts are quenched (\citealt{Cornelisse2000,Cumming2001,Kuulkers2002ks1731}).
After the envelope has cooled down, unstable burning resumes, and
Type I bursts return. Quenching times of one to several weeks are
predicted (\citealt{2004CummingMacBeth}). Because of non-contiguous
observing schedules and the presence of frequent data gaps due to
Earth occultations and South Atlantic Anomaly passages, the end of
the quenching period and the reappearance of bursts have most likely
not been observed. Only upper limits of more than one month have been
derived for the quenching time (e.g., \citealt{Kuulkers2003a}).

The return of hydrogen/helium bursts after a superburst provides a
unique opportunity to study the transition of stable to unstable burning
in the neutron star atmosphere. This transition is observed in a number
of LMXBs, and is associated with changes in the mass accretion rate
(e.g., \citealt{Cornelisse2003}). A higher rate implies a hotter
envelope as well as a faster accumulation of accreted material, which
leads to steady-state burning of hydrogen and helium. Current burst
models predict that this transition occurs at a 10 times higher accretion
rate than inferred from observations (e.g., \citealt{Heger2005}),
although increased crustal heating (e.g., \citealt{Gupta2007}) and
rotationally induced turbulent mixing (\citealt{Piro2007}) may reduce
the discrepancy (\citealt{Keek2009}). Alternatively, it may be the
result of a higher local accretion rate. 

The transition of the thermonuclear burning behavior in the atmosphere
after a superburst is solely due to the temperature change, whereas
the mass accretion rate remains constant. Note that in practice the
mass accretion rate can vary on the time scale of a superburst, which
would provide an effect additional to the change in the burning behavior.
Because the cooling after a superburst sets the thermal profile of
the envelope, this poses strict constraints on the environment in
which the first bursts ignite. Detecting the transition of stable
to unstable burning after a superburst will allow us to determine
the column depth and temperature at which the transition occurs. These
are important ingredients for improving the current models. Unfortunately,
the observations of bursts after a superburst are scarce, and at the
moment of writing the actual transition has not been observed.

In this paper we provide one-dimensional multi-zone models of a neutron
star envelope that undergoes a superburst. We create models with,
respectively, a carbon-rich, a helium-rich , and a solar-composition
atmosphere. We self-consistently simulate both the superburst and
the burning processes in the atmosphere. The burst quenching is followed,
as well as the transition to unstable burning and the return of bursts.
We focus on the nuclear reactions that are responsible for the different
phenomena in the light curves. Finally, we combine a large set of
observations to place constraints on burst quenching.

\section{Numerical method and observations}

The one-dimensional models of the neutron star envelope presented
in this paper are similar to the superburst models by \citet{Keek2011}.
We refer to that study for a detailed description of the code employed
and the setup of the models. Here we describe the main properties
of our simulations, as well as the differences with respect to the
models by \citet{Keek2011}. Most notable are the use of a different
prescription for electron conductivity and a different implementation
of accretion, which may cause small differences in the ignition conditions
of superbursts.

\subsection{Stellar Evolution Code}

We create and evolve one-dimensional models of the neutron star envelope
using the implicit hydrodynamics stellar evolution code KEPLER (\citealt{Weaver1978}).
The version of KEPLER that we use is similar to the version used in
recent studies (e.g., \citealt{Woosley2002RvMP,Woosley2004,Heger2007}).
We employ an adaptive one-dimensional Lagrangian grid in the radial
direction, under the assumption of spherical symmetry. To follow the
chemical evolution we use a large adaptive nuclear network (\citealt{Rauscher2003})
including the hot-CNO cycle (\citealt{Wallace1981}), the $3\alpha$-,
\textsl{rp}-, and \textsl{$\alpha$p}-processes, carbon fusion, and
photodisintegration. We take into account neutrino energy loss (\citealt{Itoh1996}),
as well as radiative opacity and electron conductivity (\citealt{Iben1975}).
Mixing of the chemical composition by convection, semiconvection,
and thermohaline circulation is implemented as a diffusive process
using mixing-length theory (e.g., \citealt{Clayton1968book}).

Mass accretion is implemented by increasing the pressure of the outermost
zone in the model at each time step (see also \citealt{Woosley1984,Taam1996}).
Periodically a zone is added on top. Because of the details of the
implementation, this causes a small dip in the light curve. We carefully
check that this does not influence any of the conclusions that we
draw about the bursting behavior. The light curves in this paper are
corrected by omitting the brief time intervals when the dips occur.

\subsection{Model Setup}

We model the neutron star envelope above a radius of $10\,\mathrm{km}$
and an enclosed mass of $1.4\,\mathrm{M_{\odot}}$, up to the surface.
The bottom of the surface zone is at a column depth of $y=10^{6}\,\mathrm{g\, cm^{-2}}$.
This is well below the typical column depth for hydrogen/helium bursts
of $y\simeq10^{8}\,\mathrm{g\, cm^{-2}}$. The inner part of the model
is formed by a $2\times10^{27}\,\mathrm{g}$ iron substrate. It serves
as a buffer into which heat generated by the superburst can flow,
and diffuse outward on longer time scales. This ensures a correct
late-time light curve.

Crustal heating is taken into account by a constant luminosity at
the inner boundary equivalent to $Q_{\mathrm{b}}=0.2\,\mathrm{MeV\, nucleon^{-1}}$.
Because of neutrino cooling in the substrate, the effective amount
of crustal heating at the top of the substrate, close to the superburst
ignition depth, is $Q_{\mathrm{b}}=0.1\,\mathrm{MeV\, nucleon^{-1}}$.
This is in the range of typically assumed values for superbursters
at accretion rates above $10\,\%$ of the Eddington limit (e.g., \citealt{Haensel1990,Cumming2006}).

On top of the substrate we accrete a mixture of $80\,\%$ $^{56}\mathrm{Fe}$
and $20\,\%$ $\mathrm{^{12}C}$. The latter is the typical mass fraction
of carbon that \citet{Cumming2006} found from fits to observed light
curves of hydrogen accreting superbursters, and $^{56}\mathrm{Fe}$
is the most abundant isotope in the ashes of hydrogen-rich bursts
(e.g., \citealt{Woosley2004}). The ashes of helium-rich bursts consist
mostly of somewhat lighter isotopes: $^{28}\mathrm{Si}$, $^{32}\mathrm{S}$,
$^{36}\mathrm{Ar}$, $^{40}\mathrm{Ca}$ (e.g., \citealt{Joss1980}).
For our models, however, we use $^{56}\mathrm{Fe}$, in order to have
a consistent opacity of all envelope models, giving comparable ignition
conditions for the superburst.

Accretion of a hydrogen or helium-rich atmosphere is done only briefly
before the superburst, because the computational expense of simulating
the many short bursts that occur before the superburst is prohibitive.
From a model where only a carbon-rich mixture was accreted we know
already the moment of superburst ignition. $12.7\,\mathrm{hr}$ before
this time, we replace the accretion composition by a solar ($71\,\%$
by mass $^{1}$H, $27\,\%$ $^{4}$He, $2\,\%$ $^{14}$N), or a helium-rich
mixture ($98\,\%$ $^{4}$He, $2\,\%$ $^{14}$N). This allows for
several hydrogen and/or helium bursts to take place before the superburst,
ensuring equilibrium is reached for the effects of chemical inertia
(\citealt{Woosley2004}). The change of accretion composition does
not affect the ignition of the superburst, because it depends on the
pressure at the bottom of the carbon layer, which continues to increase
at the some rate. Moreover, at that time the initial phase of the
carbon runaway has already started.

An accretion rate of $\dot{M}=5.25\times10^{-9}\,\mathrm{M_{\odot}}\,\mathrm{yr^{-1}}$
is used. For an atmosphere of solar composition on a neutron star
of $1.4\,\mathrm{M_{\odot}}$ this corresponds to $30\,\%$ of the
Eddington-limited rate $\dot{M}_{\mathrm{Edd}}=1.75\times10^{-8}\,\mathrm{M_{\odot}}\,\mathrm{yr^{-1}}$. 

The presented results are not corrected for the redshift due to the
neutron star's gravity (see also \citealt{Keek2011}). Our Newtonian
model has the same surface gravity as when general relativity is taken
into account for a star with the same mass and a $11.2\,\mathrm{km}$
radius, which has a gravitational redshift of $z=0.26$ (e.g., \citealt{Woosley2004}).

Based on the atmosphere composition, we refer to the simulations as
model ``C'', ``He'', and ``H'' (Table~\ref{tab:Superburst-properties-for}).

\subsection{MINBAR Catalog of Observations}

To derive observational constraints on the bursting behavior of superbursters
we employ version 0.51 of the Multi-INstrument Burst ARchive (MINBAR;
\citealt{Keek2010}; see \href{http://users.monash.edu.au/~dgallow/minbar}{http://users.monash.edu.au/$\sim$dgallow/minbar}
for more details). This catalog contains the results of the analysis
of 4,192 observed Type I X-ray bursts from 72 sources as well as 27,340
pointings on 84 sources. The observations have been performed with
the Wide-Field Cameras (WFCs) on board the \emph{Beppo Satellite per
Astronomia X} (\emph{BeppoSAX}; \citealt{Cornelisse2003}) and the
Proportional Counter Array (PCA) on board the \emph{Rossi X-Ray Timing
Explorer} (\emph{RXTE}; \citealt{Galloway2008catalog}). Both instruments
are sensitive in a similar energy range above $2\,\mathrm{keV}$.
Because of its larger collecting area, the PCA is more sensitive to
faint bursts (e.g., \citealt{Keek2010}).

MINBAR comprises the largest collection of X-ray bursts available,
but we refer to the literature if more constraining observations have
been reported.

\section{Results}

\subsection{Hydrogen/Helium Atmosphere Models}

\begin{figure*}
\includegraphics{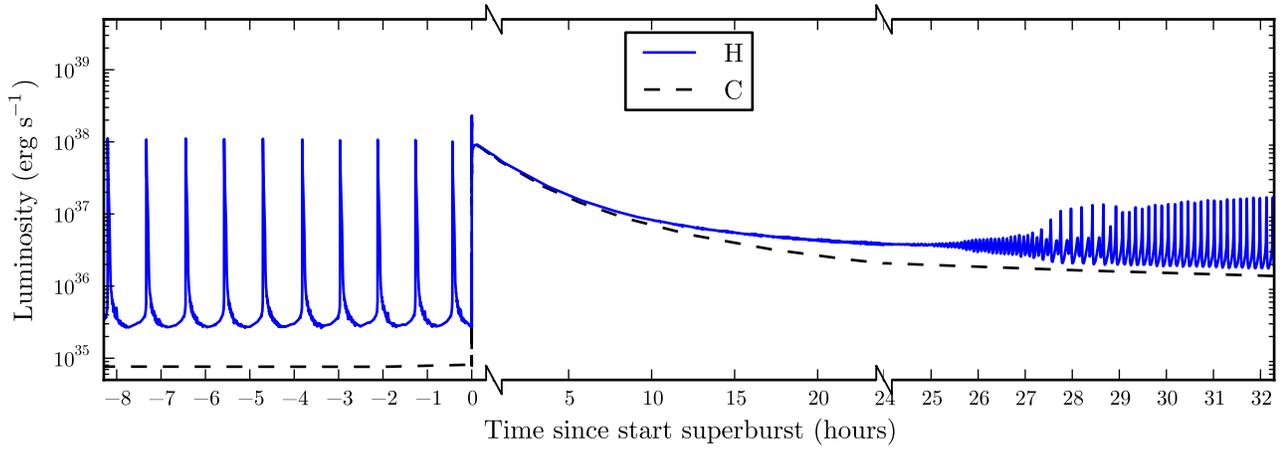}

\caption{\label{fig:longlc_h}Light curve of models H and C. Breaks on the
axis indicate a change in time scale. Only a small fraction of the
simulated bursts after the superburst are shown.}
\end{figure*}

\begin{figure*}
\includegraphics{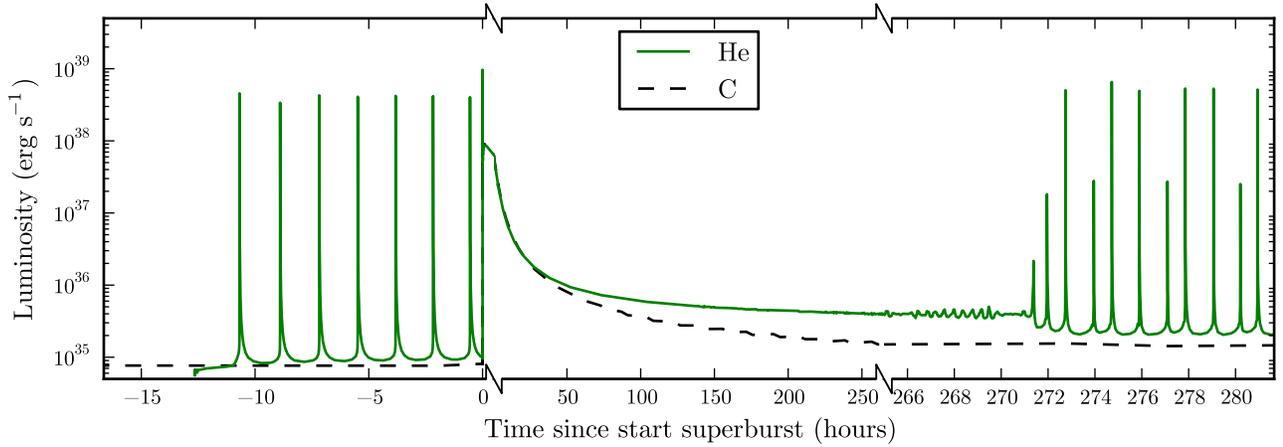}

\caption{\label{fig:longlc_he}Light curve of models He and C. Breaks on the
axis indicate a change in time scale. Only a small fraction of the
simulated bursts after the superburst are shown.}
\end{figure*}

Initially the accretion of carbon-rich material is simulated to self-consistently
build up a neutron star envelope close to the ignition of a superburst.
Approximately half a day before the thermonuclear runaway of carbon
burning sets in, the accretion composition is changed. We create three
different models: one accreting solar composition (model H), one with
helium-rich material (model He), and for comparison one where we retain
the carbon-rich atmosphere (model C). From that moment we follow both
the carbon burning in the ocean and the hydrogen/helium burning in
the atmosphere. Several normal X-ray bursts occur before the superburst
(Fig.~\ref{fig:longlc_h}, \ref{fig:longlc_he}). During the superburst
decay, hydrogen and helium burn stably: bursts are quenched. Once
the envelope has cooled sufficiently, unstable burning resumes. At
the transition, oscillatory burning takes place (marginally stable
burning). The first bursts after the superburst are less energetic
and have shorter recurrence times than the bursts before the superburst.
When the envelope cools down further, the recurrence times become
longer and the bursts become as energetic as they were before the
superburst. For the models H and He we calculate over 900 bursts per
model.

\subsection{Superburst}

\begin{figure*}
\includegraphics{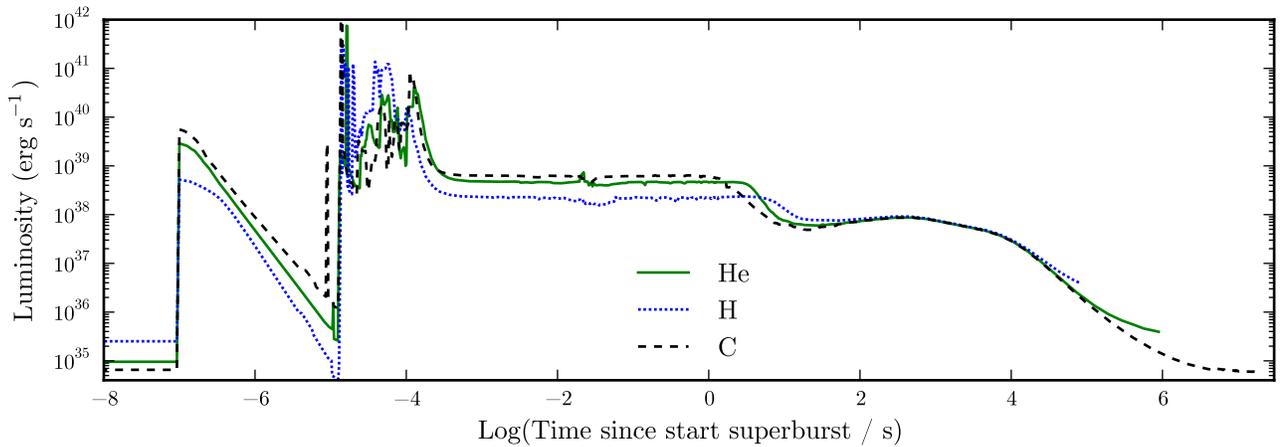}

\caption{\label{fig:log_sb}Logarithmic light curve of superbursts from models
He, H, and C. The curves end when the burst-quenching period ends,
or, in the case of a carbon-rich atmosphere, when the next superburst
occurs.}
\end{figure*}
The superburst occurs $1.28\,\mathrm{years}$ after the start of the
simulations, after accreting a column of $y=1.1\times10^{12}\,\mathrm{g\, cm^{-2}}$.
Its ignition is not in-phase with the occurrence of hydrogen/helium
bursts (Fig.~\ref{fig:longlc_h}, \ref{fig:longlc_he}): flashes
in the envelope do not trigger the runaway carbon burning. The superburst
occurs slightly earlier in the models H and He compared to C: 1.1~minutes
for helium accretion and 5.2~minutes for solar composition. This
may be due to heating from hydrogen and helium burning, but the differences
in recurrence time, and therefore ignition depth, are too small to
cause substantial differences in, for example, the superburst fluence
or other observables. The amount of time before the superburst during
which we simulate the hydrogen/helium bursts may not have been long
enough to bring the model into thermal equilibrium, and the deeper
layers could still be increasing somewhat in temperature because of
heating by the bursts. Because most of the energy produced by the
bursts is radiated away from the surface, however, this introduces
only a minor deviation in the temperature at the carbon ignition depth,
which can be modeled by a slightly higher effective $Q_{\mathrm{b}}$.

The superbursts start with a detonation, which drives a shock toward
the surface, and produces a brief shock breakout peak in the light
curve (Fig.~\ref{fig:log_sb}), followed on a dynamical time-scale
by a precursor burst. The superburst peak is reached $6.4$~minutes
after the superburst onset. Stable hydrogen (helium) burning produces
a peak luminosity, $L_{\mathrm{peak}}$, that is $5\,\%$ ($0.4\,\%$)
higher for model H (He) compared to C (Table~\ref{tab:Superburst-properties-for}).
The decays follow the same two-component power-law profile, with again
hydrogen burning raising the luminosity by a few percent. The light
curves for the models H and C deviate from the cooling curve of C
when the luminosity becomes comparable to that from stable hydrogen
or helium burning. The hydrogen or helium burning contribution to
the total luminosity exceeds $10\,\%$ after roughly $\sim2\times10^{4}\,\mathrm{s}$
for model H, and after $\sim1\times10^{5}\,\mathrm{s}$ for He. For
all models a small amount of carbon burns at the bottom of the newly
accreted material. The burning time scale, however, is much longer
than the accretion time, so this is not steady-state burning, but
leads up to the next superburst.

\begin{table}

\caption{\label{tab:Superburst-properties-for}Properties of Superbursts and
Subsequent Bursts for Different Atmosphere Compositions}

\begin{centering}
\begin{tabular}{lccc}
\hline 
Model & C & He & H\tabularnewline
\hline 
Accretion composition & $0.2$ $^{12}$C & $0.98$ $^{4}$He & $0.71$ $^{1}$H\tabularnewline
(mass fractions) & $0.8$ $^{56}$Fe & $0.02$ $^{14}$N & $0.27$ $^{4}$He\tabularnewline
 &  &  & $0.02$ $^{14}$N\tabularnewline
$L_{\mathrm{peak}}\,(10^{37}\,\mathrm{erg\, s^{-1}})$ & $8.7$ & $8.8$ & $9.3$\tabularnewline
$E_{\mathrm{shock\, br}}\,(10^{32}\,\mathrm{erg})$ & $7.7$ & $4.4$ & $1.4$\tabularnewline
$L_{\mathrm{shock\, br}}\,(10^{39}\,\mathrm{erg\, s^{-1}})$ & $10$ & $6.2$ & $1.3$\tabularnewline
Precursor duration (s) & $1.2$ & $3.2$ & $5.7$\tabularnewline
$E_{\mathrm{precursor}}\,(10^{39}\,\mathrm{erg})$ & $0.75$ & $1.5$ & $1.3$\tabularnewline
$t_{\mathrm{minimum}}$ (s) $\mathrm{^{a}}$ & $20$ & $25$ & $42$\tabularnewline
$t_{\mathrm{quench}}$ (days) &  & $11.3$ & $1.1$\tabularnewline
$P_{\mathrm{osc}}$ (minutes) $\mathrm{^{b}}$ &  & $20$ & $5.0$\tabularnewline
$t_{\mathrm{return}}$ (days) $\mathrm{^{c}}$ &  & $115$ & $35$\tabularnewline
\hline 
\end{tabular}
\par\end{centering}

$\mathrm{^{a}}$ Time of minimum luminosity between precursor and
peak.

$\mathrm{^{b}}$ Period of oscillatory burning.

$\mathrm{^{c}}$ Time when burst properties return to pre-superburst
values.

\end{table}

\subsection{Shock Breakout}

After the superburst detonation, a shock travels outward from the
bottom of the carbon-rich layer. Once it reaches the surface, a shock
breakout peak is produced in the light curve (Fig.~\ref{fig:log_sb}).
We determine its maximum luminosity, $L_{\mathrm{shock\, br}}$, and
its fluence, $E_{\mathrm{shock\, br}}$, as measured within $5\times10^{-6}\,\mathrm{s}$
after the onset (Table~\ref{tab:Superburst-properties-for}). The
differences in $L_{\mathrm{shock\, br}}$ trace variations in the
opacity of the outer atmosphere.

The shock accelerates the outer zones, where the density is lowest,
and the shock over-pressure is highest, to a substantial fraction
of the speed of light within a very short time interval. This behavior
is likely not well resolved by our model, and at times introduces
large variability in, for example, the luminosity in the outer zones
during the shock breakout and the start of the subsequent precursor
burst. During this part of the superburst we take the mean luminosity
of the 10 outermost zones to reduce the effect on the light curve,
although some of the introduced variability remains visible (Fig.~\ref{fig:log_sb}).

\subsection{Precursor}

\begin{figure}
\includegraphics[angle=90,width=1\columnwidth]{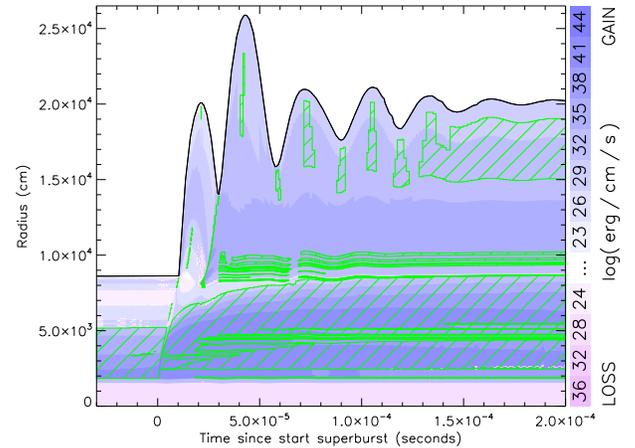}

\caption{\label{fig:h_precursor}For model He: energy generation/loss (color
scale) in the neutron star envelope as a function of time in a short
interval around the superburst onset. Green hatching indicates convection
(Rayleigh--Taylor instabilities).}
\end{figure}

\begin{figure}
\includegraphics[angle=90,width=1\columnwidth]{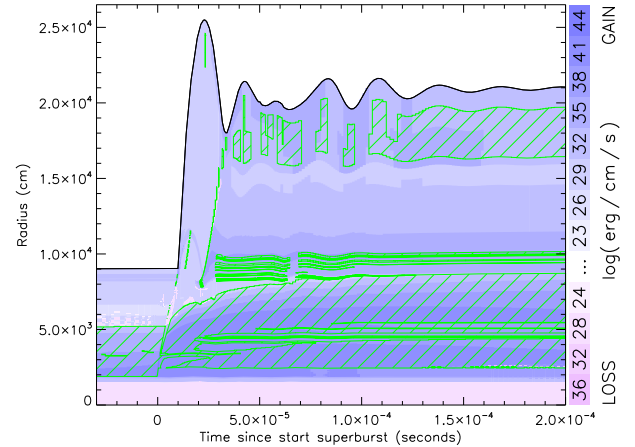}

\caption{\label{fig:he_precursor}Same as Fig.~\ref{fig:h_precursor} for
model H.}
\end{figure}

Most of the energy of the shock is used to expand the outer layers.
They fall back on a dynamical time scale of $10^{-5}\,\mathrm{s}$,
and dissipate that energy into heat, which powers a precursor burst
that reaches the Eddington limit. At the start of the precursor burst,
the fallback causes the material to settle while undergoing a damped
oscillation. This creates corresponding variability of super-Eddington
luminosity as high as $3.7\times10^{40}\,\mathrm{erg}\mathrm{\, s^{-1}}$
(Fig.~\ref{fig:h_precursor}, \ref{fig:he_precursor}).

The precursor light curves are dependent on the atmospheric composition
(Fig.~\ref{fig:log_sb}). The heating of the atmosphere instigates
the burning of hydrogen and/or helium upon fallback (Fig.~\ref{fig:h_precursor},
\ref{fig:he_precursor}), leading to a thermonuclear runaway. This
adds to the precursor fluence $E_{\mathrm{precursor}}$ (Table~\ref{tab:Superburst-properties-for}).
In model H $26\,\%$ of the $7.4\times10^{7}\,\mathrm{g}$ column
of solar composition is burned, and in model He $67\,\%$ of the $9.0\times10^{7}\,\mathrm{g}$
helium column. Because of the difference in energy yield of the hydrogen
and helium burning nuclear reactions, the total energy released and,
hence, the fluence is very close for these particular compositions.

The luminosity reaches a plateau caused by photospheric radius expansion
(PRE, Fig.~\ref{fig:log_sb}). We compute the duration of the PRE
phase as the time from the precursor onset to the time when the luminosity
drops below $90\%$ of the plateau value (Table~\ref{tab:Superburst-properties-for}).

After the PRE phase, the luminosity drops quickly, reaching a minimum
at $t_{\mathrm{minimum}}$ (Table~\ref{tab:Superburst-properties-for}),
and subsequently climbing to the superburst peak at approximately
$400\,\mathrm{s}$.

The precursor light curve contains some irregularities. In the figures
we have filtered out strong variations occurring from one time step
to the next, that are due to the outer few grid points, and not representative
of the overall behavior of the model.

\subsection{Bursts in a Helium-rich Atmosphere}

\begin{figure*}
\begin{centering}
\includegraphics{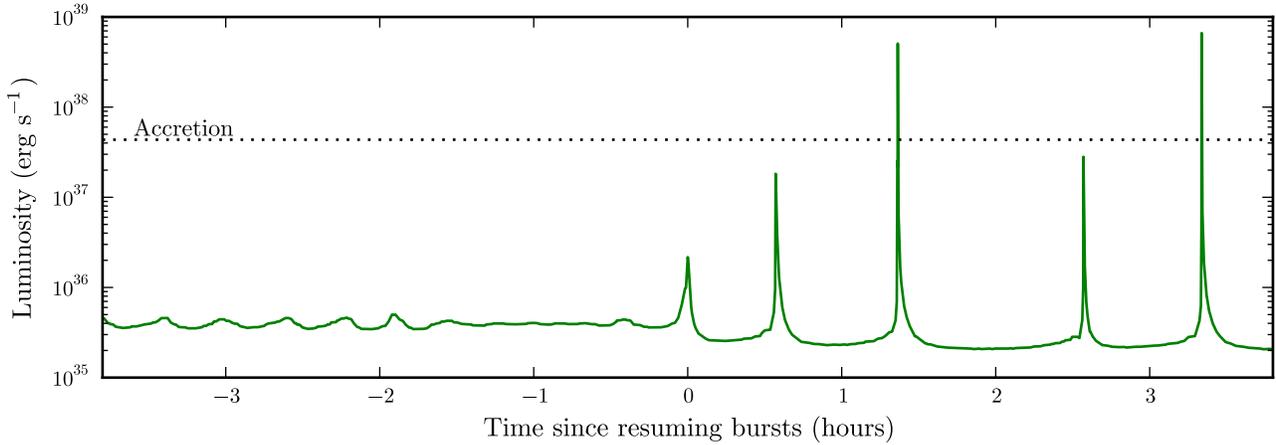}
\par\end{centering}

\caption{\label{fig:resume_he}Light curve of model He at the time when quenching
stops and bursting resumes. The dotted line indicates the level of
the accretion flux.}
\end{figure*}

First we discuss the effect of the superburst on the thermonuclear
processes in model He. After the superburst precursor, helium burning
in the atmosphere proceeds in a stable manner. Bursts only reappear
once the envelope has cooled down sufficiently after $11.3\,\mathrm{days}$
(Fig.~\ref{fig:longlc_he}).

Before bursts resume, the burning is marginally stable when the temperature
in the atmosphere drops to $3\times10^{8}\,\mathrm{K}$, leading to
oscillations in the light curve with a period close to $20\,\mathrm{minutes}$
(Fig.~\ref{fig:resume_he}). The oscillations have a small amplitude
of at most approximately $10\,\%$ of the average luminosity, and
numerical noise dampens the oscillations during certain periods. 
\begin{figure}
\includegraphics[bb=183bp 250bp 428bp 534bp,clip]{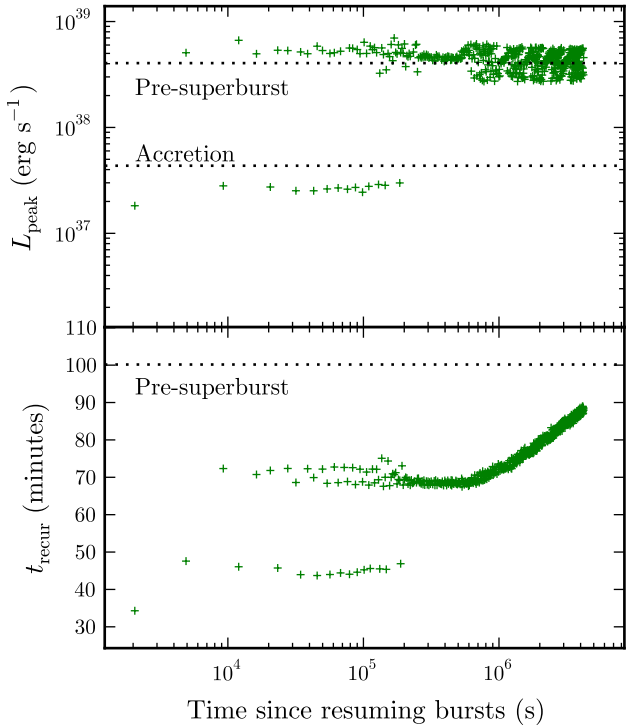}

\caption{\label{fig:trecur_20}Peak luminosity $L_{\mathrm{peak}}$ and recurrence
time $t_{\mathrm{recur}}$ of bursts after the quenching period for
model He. The dotted lines indicate the mean values for bursts before
the superburst and the luminosity of the accretion process.}
\end{figure}
Once the bursts start, the burst peak flux increases, and the third
burst already has a peak luminosity comparable to the bursts before
the superburst (Fig.~\ref{fig:resume_he}, \ref{fig:trecur_20}).
During $2.3$~days we find both bright bursts with a peak luminosity
of $L_{\mathrm{peak}}\simeq4\times10^{38}\,\mathrm{erg\, s^{-1}}$
(the Eddington luminosity for a hydrogen-deficient atmosphere) and
weaker bursts with $L_{\mathrm{peak}}\simeq3\times10^{37}\,\mathrm{erg\, s^{-1}}$.
The bright as well as the weak bursts have a relatively slow rise
of the light curve. During this time $3\alpha$ is the predominant
nuclear process, producing $^{12}$C. In the bright bursts $3\alpha$
raises the temperature sufficiently for $\alpha$-captures to take
over, causing a faster rise of the luminosity, and producing mostly
$^{28}$Si. The bright bursts heat the envelope sufficiently for additional
stable helium burning to take place, followed by a weak burst. Stable
helium burning reduces the helium content of the atmosphere, leading
to a burst with a lower $L_{\mathrm{peak}}$. The weak burst does
not heat the envelope enough for additional stable burning, such that
the next burst is again bright, and ignites after a shorter recurrence
time. While the envelope continues to cool from the superburst, the
number of bright bursts in between weak bursts increases, until after
$2.3$~days the weak bursts disappear.

\begin{figure}
\includegraphics{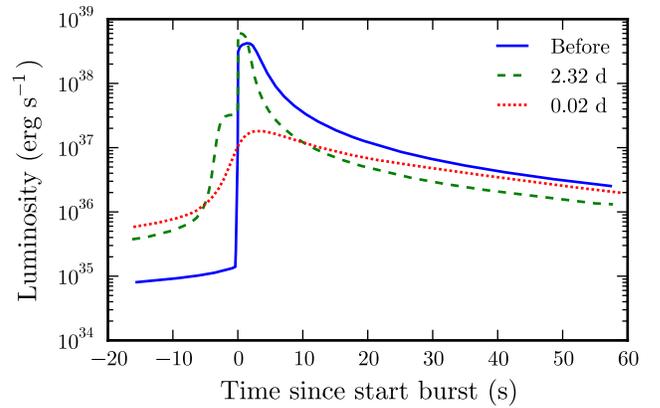}

\caption{\label{fig:he_bursts}Comparison of light curves of helium bursts
aligned on the peak. The smallest burst occurred 34~minutes after
bursts resumed, and the one with the bump before the peak occurred
after 2.3~days. The burst with the longest decay occurred before
the superburst. Superburst emission has been subtracted for all bursts.}
\end{figure}

Comparing the profiles of individual bursts after the quenching period,
the weak bursts have a relatively long rise of $\sim5\,\mathrm{s}$
(Fig.~\ref{fig:he_bursts}). The early bright bursts share this slow
rise to a similar luminosity as the peak of the weak bursts, but then
transition in a fast sub-second rise to the Eddington luminosity.
Later bright bursts show an initial ``bump'' that decreases in duration
over time. The bursts from before the superburst show a slow rise
component with a duration of $0.4\,\mathrm{s}$. The decay of the
bursts becomes longer as the recurrence time increases, reflecting
the longer thermal time scales of increasing ignition depths.

The flux from the accretion process at a rate of $\dot{M}=5.25\times10^{-9}\,\mathrm{M_{\odot}}\,\mathrm{yr^{-1}}$
for a neutron star of $1.4\,\mathrm{M_{\odot}}$ with a $10\,\mathrm{km}$
radius is $4.35\times10^{37}\,\mathrm{erg\, s^{-1}}$ (assuming isotropic
emission and a $100\,\%$ efficient accretion process). The initial
oscillations and weak bursts have a lower $L_{\mathrm{peak}}$, but
the brighter bursts outshine the accretion flux by approximately a
factor $10$.

The superburst burns most carbon out to $y\simeq7\times10^{8}\,\mathrm{g\, cm^{-2}}$.
Between this depth and the ignition depth of the helium bursts at
$y\simeq1\times10^{8}\,\mathrm{g\, cm^{-2}}$, a carbon mass fraction
of $0.1$ survives the runaway burning, but burns on a longer time
scale. Carbon burns through $^{12}\mathrm{C}(^{12}\mathrm{C},\alpha)^{20}\mathrm{Ne}$
and subsequent $\alpha$-capture reactions, producing predominantly
$^{28}$Si. When we stop the simulation, the carbon mass fraction
of this material varies from $2\times10^{-3}$ down to $2\times10^{-4}$.

\begin{figure}
\includegraphics{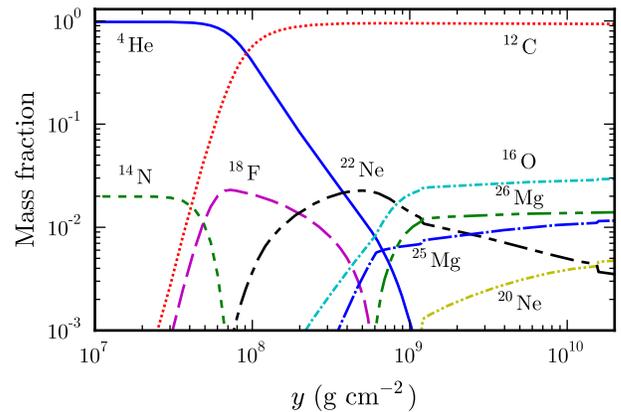}

\caption{\label{fig:he_quench}Composition of the envelope at the end of the
burst quenching period, down to the bottom of the layer of ashes from
stable helium burning. Only the most abundant isotopes are shown.}
\end{figure}

Directly after the superburst, during the burst quenching period,
helium burns stably to carbon by the $3\alpha$-process. Captures
of $\alpha$ on $^{12}$C and $^{14}$N from the accretion composition
produce a limited fraction of $^{16}$O, $^{18}$F, and heavier isotopes
up to magnesium (Fig.~\ref{fig:he_quench}). Over time, as the atmosphere
cools down, the $\alpha$-capture rates reduce, and only lighter elements
up to neon are produced. The main product of nuclear burning during
the quenching period, however, is $^{12}$C, with a mass fraction
of $95\,\%$. After the quenching period, this material is compressed
to higher densities, and $^{12}\mathrm{C}(^{12}\mathrm{C},\alpha)^{20}\mathrm{Ne}$
reduces the carbon mass fraction. Subsequent $\alpha$-capture reactions
produce mainly $^{28}$Si, $^{32}$S, and $^{36}$Ar. The next superburst
ignition occurs close to the bottom of this layer, and depends on
the remaining carbon fraction. When we end the simulation, the $^{12}$C
fraction is $23\,\%$, and still dropping.

During the quenching period, most burning takes place close to $y\simeq10^{8}\,\mathrm{g\, cm^{-2}}$,
but not all helium is burned. Some survives down to a depth of $y\simeq10^{10}\,\mathrm{g\, cm^{-2}}$.
The low mass fraction ($Y\lesssim10^{-3}$) makes $3\alpha$ inefficient,
but series of $\alpha$-captures ($^{12}\mathrm{C}(\alpha,\gamma){}^{16}\mathrm{O}(\alpha,\gamma){}^{20}\mathrm{Ne}(\alpha,\gamma)^{24}\mathrm{Mg}$)
still occur. After stable burning at $y\simeq10^{8}\,\mathrm{g\, cm^{-2}}$
ends, the $\alpha$-captures in the deeper layers continue until helium
is fully depleted. During this time, even though the helium mass fraction
is low, because of the large total mass down to $y\simeq10^{10}\,\mathrm{g\, cm^{-2}}$,
the captures contribute a substantial fraction of the total generated
energy at any given time. This effectively slows down the cooling
for up to $8.1\,\mathrm{days}$, keeping the recurrence time of the
bursts constant. Once helium is depleted at these depths, the cooling
continues, and the burst recurrence time increases over time. $46.3$~days
after burst resumption the recurrence time is $88$~minutes, whereas
the pre-superburst recurrence time was $100$~minutes. We stop the
simulation here, but extrapolating the trend, the original recurrence
time will be recovered approximately $115$~days after burst resumption.

The model produces bursts with a spread of $\sim50\,\%$ in $L_{\mathrm{peak}}$
(Fig.~\ref{fig:trecur_20}) with a quasi-periodic behavior. To a
lesser extend these variations can also be seen in $t_{\mathrm{recur}}$.
This is caused by the dependence of the ignition conditions of a burst
on the previous bursts, based on, e.g., the fraction of helium that
was burned, the heat deposition, and compositional inertia.

\subsection{Bursts in a Solar-composition Atmosphere}

\begin{figure*}
\begin{centering}
\includegraphics{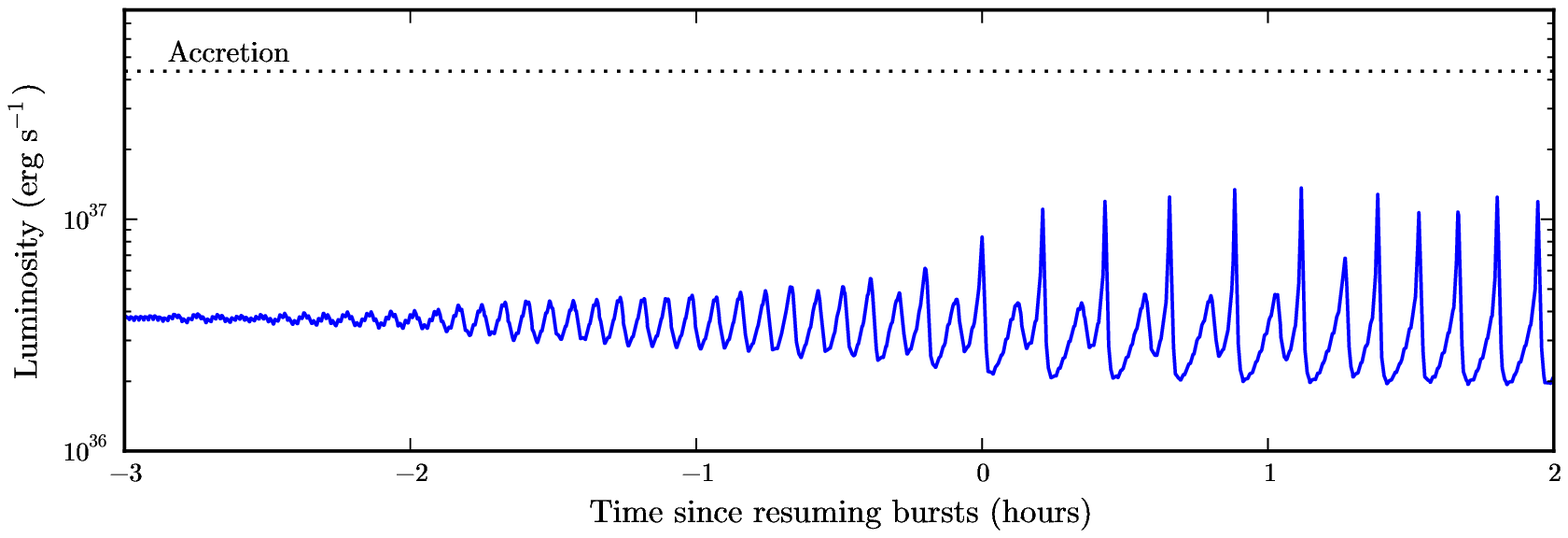}
\par\end{centering}

\caption{\label{fig:resume_h}Light curve of model H at the time when quenching
stops and bursting resumes. The dotted line indicates the level of
the accretion flux.}
\end{figure*}

\begin{figure}
\includegraphics{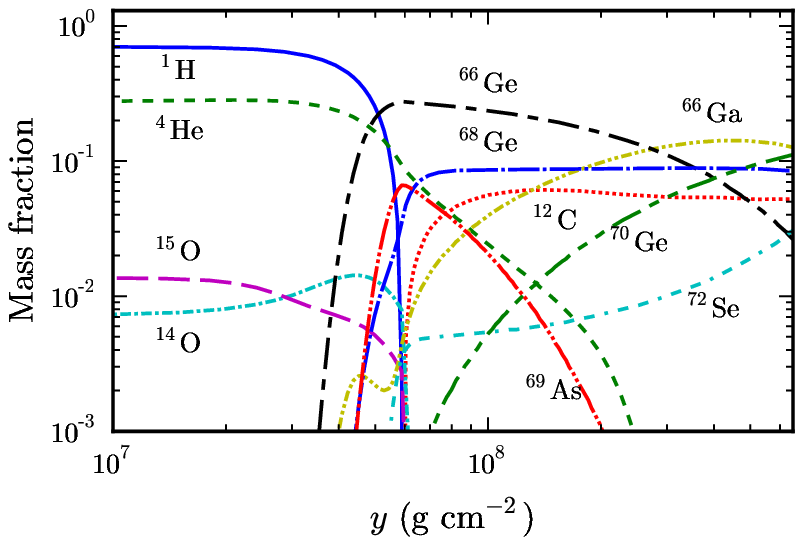}

\caption{\label{fig:h_quench}Composition of the envelope at the end of the
burst-quenching period, down to the bottom of the layer of ashes from
stable burning of solar-composition material. Only a selection of
the most abundant isotopes is shown.}
\end{figure}

Burst quenching is also present in model H (Fig.~\ref{fig:longlc_h},
\ref{fig:resume_h}). Directly following the superburst, hydrogen
and helium burning is stable. Helium burns by the $3\alpha$-process,
providing CNO elements which facilitate hydrogen burning by the hot-CNO
cycle. Break-out from this cycle by $^{15}\mathrm{O}(\alpha,\gamma)^{19}\mathrm{Ne}$
allows for a rapid series of proton captures and $\beta$-decays (\textsl{rp}-process)
to produce mostly $^{66}\mathrm{Ga}$ and $^{66}\mathrm{Ge}$, whereas
a large number of isotopes with mass numbers in the range of $59$--$72$
contribute mass fractions of several percents (Fig.~\ref{fig:h_quench}).

Similarly to model He, after the superburst a layer with a $^{12}$C
mass fraction of $10\,\%$ remains between $y\simeq1\times10^{8}\,\mathrm{g\, cm^{-2}}$
and $y\simeq7\times10^{8}\,\mathrm{g\, cm^{-2}}$. It burns on a longer
time scale down to a mass fraction of $2\times10^{-4}$. Hydrogen
and helium burning during the burst quenching period produce carbon
at a mass fraction of $5\,\%$. Nuclear burning on longer time scales
reduces the carbon mass fraction in this layer to $10^{-3}$.

Burst quenching is much shorter than in the model with a helium-rich
atmosphere: bursts resume after $1.1\,\mathrm{days}$. As the superburst
cools, the temperature drops below $4.3\times10^{8}\,\mathrm{K}$,
and the breakout reaction $^{15}\mathrm{O}(\alpha,\gamma)^{19}\mathrm{Ne}$
becomes less efficient than the $3\alpha$-process. Whereas at higher
temperatures the break-out reactions quickly removed CNO elements,
now the CNO mass fraction is growing. This causes an increase in the
helium production through the CNO cycle, and results in the increase
of the energy generation rate of both the CNO cycle and the $3\alpha$-process.
The involved reactions raise the atmosphere temperature until once
more the $^{15}\mathrm{O}(\alpha,\gamma)^{19}\mathrm{Ne}$ break-out
can efficiently remove most of the CNO elements, providing seed nuclei
for the \textsl{rp}-process, which captures most of the hydrogen.
With hydrogen and the CNO isotopes gone, the CNO cycle and rp-process
switch off, reducing the helium production, and thereby the $3\alpha$
rate, allowing the atmosphere to cool down. Note that while hydrogen
is depleted, helium is not. As fresh hydrogen and helium are accreted
and mixed in from layers closer to the surface, the cycle repeats
itself. This produces a series of oscillations in the light curve,
that announces the end of the steady-state burning of the burst quenching
phase (Fig.~\ref{fig:resume_h}). The time scale for the oscillations
is $5.0\,\mathrm{min}$. 

\begin{figure}
\includegraphics{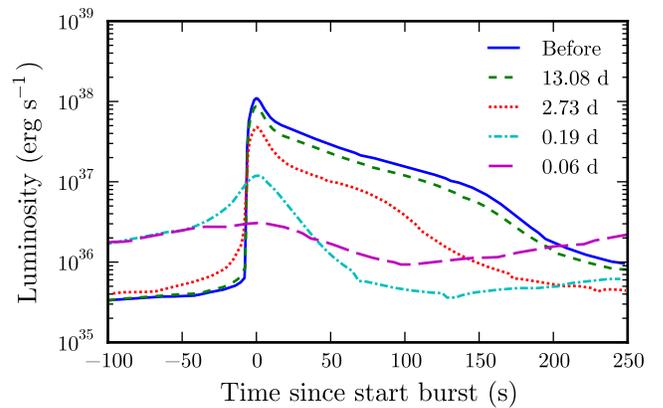}

\caption{\label{fig:h_bursts}Comparison of light curves of hydrogen/helium
bursts aligned on the peak. From small peak to high, the bursts occurred
1.5~hr, 4.6~hr, 2.7~days, and 13.1~days after bursting resumed,
respectively. The brightest burst occurred before the superburst.
Superburst luminosity has been subtracted for all bursts.}
\end{figure}

The oscillations increase in amplitude over time. After $2.1\,\mathrm{hr}$
the $^{15}\mathrm{O}(\alpha,\gamma)^{19}\mathrm{Ne}$ breakout and
subsequent \textsl{rp}-process increase the temperature during the
oscillations sufficiently for the $^{14}\mathrm{O}(\alpha,p)^{17}\mathrm{F}(p,\gamma)^{18}\mathrm{Ne}$
breakout reactions initiate the \textsl{$\alpha$p}-process, which
causes a faster rise of the luminosity, producing small bursts instead
of oscillations. For $\sim1.5\,\mathrm{hr}$ oscillations and small
bursts alternate, until an equilibrium is reached and only small bursts
occur (Fig.~\ref{fig:resume_h}). The bursts following an oscillation
have a shorter recurrence time. Initially the bursts have a slow rise,
similar to the oscillations, and a faster decay. Over time, the rise
shortens, and the decay lengthens (Fig.~\ref{fig:h_bursts}). 

When the oscillations have disappeared, and only bursts remain, the
recurrence time is constant at $8.2\,\mathrm{minutes}$ until approximately
$1.2\,\mathrm{days}$ after burst resumption. As in model He, this
is due to the burning of residual helium at greater depths, which
delays the cooling of the atmosphere. Although the recurrence time
is constant, the peak luminosity $L_{\mathrm{peak}}$ increases with
time.

\begin{figure}
\includegraphics[bb=183bp 250bp 428bp 534bp,clip]{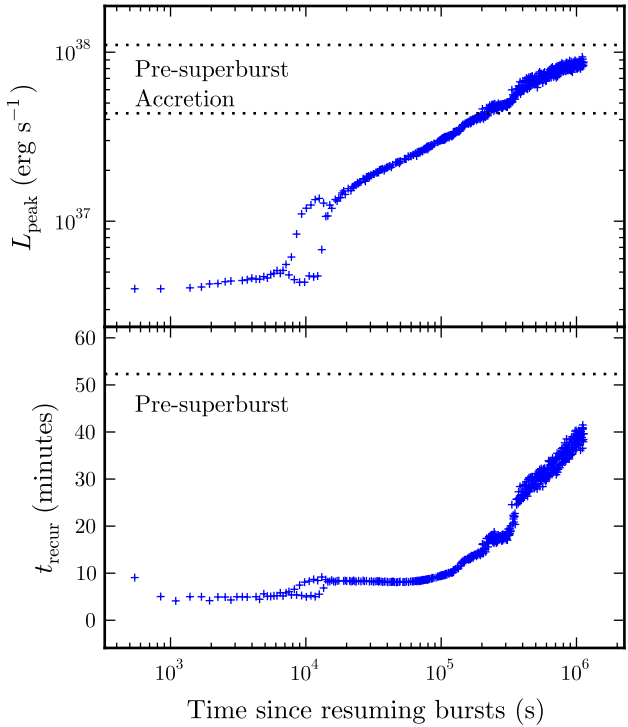}

\caption{\label{fig:trecur_19}Peak luminosity $L_{\mathrm{peak}}$ and recurrence
time $t_{\mathrm{recur}}$ of bursts after the quenching period for
model H. The dotted lines indicate the mean values for bursts before
the superburst and the luminosity of the accretion process.}
\end{figure}

Afterward both the recurrence time and $L_{\mathrm{peak}}$ increase
with time (Fig.~\ref{fig:trecur_19}). After 13~days $t_{\mathrm{recur}}=39\,\mathrm{minutes}$
is reached. The burst peak luminosity is then at $77\,\%$ of the
pre-superburst level. The burst light curve approaches the profiles
of the bursts before the superburst. At this time we end the simulation.
Extrapolating, the recurrence time of $52\,\mathrm{minutes}$ of the
bursts before the superburst will be reached approximately $35$~days
after burst resumption. 

For $1\times10^{5}\lesssim t\lesssim4\times10^{5}\,\mathrm{s}$ there
are some `oscillatory' variations in both $t_{\mathrm{recur}}$ and
$L_{\mathrm{peak}}$, which are likely of similar origin as the late-time
variations in the burst properties in the helium-rich atmosphere (Fig.~\ref{fig:trecur_19}).

\subsection{Observational Limits on Burst Quenching}

The first normal burst observed from a source after a superburst provides
an upper limit to the burst quenching time. For all known (candidate)
superbursts, we identify the first detected burst either from MINBAR,
or from the literature (Table~\ref{tab:Observational-limits-on}).
The superbursts from \object{4U~1735--444} and \object{4U~1820--303}
occurred when (or close to when) the source exhibited a persistent
flux where no or very few bursts have been detected for the respective
sources. Most probably hydrogen and helium were undergoing steady-state
burning already before the superburst started. For this reason we
exclude these superbursts when we investigate constraints on burst
quenching. Furthermore, we do not consider the recent superbursts
from \object{EXO~1745--248}, \object{SAX~J1747.0--2853}, and
\object{SAX~J1828.5--1037}, which are still being analyzed at the
moment of writing, nor those from GX~17+2, which has an atypically
high mass accretion rate (\citealt{Zand2004}).

\begin{table}
\caption{\label{tab:Observational-limits-on}Observational Limits on Burst
Quenching for All Known (Candidate) Superbursts}

\begin{centering}
\begin{tabular}{ll>{\raggedleft}p{1.5cm}r}
\hline 
Source & Time (MJD)$\mathrm{^{a}}$ & First burst (d)$^{\mathrm{b}}$ & $t_{\mathrm{exp}}$ (hr)$\mathrm{^{c}}$\tabularnewline
\hline 
\object{4U 0614+091} & 53441.70 (-0.26) & 18.6$\mathrm{^{d}}$ & 0.9$^{\,\,}$\tabularnewline
\object{4U 1254--690} & 51187.39 & 124.7$^{\,\,}$ & 25.5$^{\,\,}$\tabularnewline
\object{4U 1608--522}$\mathrm{^{t}}$ & 53495.08 & 99.8$\mathrm{^{e}}$ & 2.6$^{\,\,}$\tabularnewline
\object{4U 1636--536} & 50253.61 (-0.07) & 96.3$^{\,\,}$ & 0.0$^{\,\,}$\tabularnewline
 & 50642.37 (-0.07) & 68.6$^{\,\,}$ & 2.1$^{\,\,}$\tabularnewline
 & 51324.21 (-0.06) & 15.0$^{\,\,}$ & 33.9$^{\,\,}$\tabularnewline
 & 51962.70 & 22.8$^{\,\,}$ & 9.7$^{\,\,}$\tabularnewline
\object{KS 1731--260} & 50349.42 & 34.4$^{\,\,}$ & 48.0$^{\,\,}$\tabularnewline
\object{4U 1735--444} & 50318.13 (-0.02) & 374.3$^{*}$ & 130.6$^{\,\,}$\tabularnewline
\object{EXO 1745--248}$\mathrm{^{t}}$ & 55858.53 (-0.06) & -$^{\,\,}$ & -$^{\,\,}$\tabularnewline
\object{GX 3+1} & 50973.04 (-0.08) & 94.2$^{\,\,}$ & 2.5$^{\,\,}$\tabularnewline
\object{SAX J1747.0--2853}$\mathrm{^{t}}$ & 55605.54 & 25.1$\mathrm{^{f}}$ & -$^{\,\,}$\tabularnewline
\object{GX 17+2} & 50340.30 (-0.03) & 2.2$^{\,\,}$ & 36.6$^{\,\,}$\tabularnewline
 & 51444.10 (-0.04) & -$\mathrm{^{g}}$ & 132.0$^{\,\,}$\tabularnewline
 & 51452.33 & 2.3$^{\,\,}$ & 113.4$^{\,\,}$\tabularnewline
 & 51795.34 & 12.8$^{\,\,}$ & 100.5$^{\,\,}$\tabularnewline
\object{4U 1820--303} & 51430.07 & 167.0 & 113.7$^{\,\,}$\tabularnewline
 & 55272.72 (-0.06) & -$^{*}$ & -$^{\,\,}$\tabularnewline
\object{SAX J1828.5--1037}$\mathrm{^{t}}$ & 55877.34 (-0.06) & -$^{\,\,}$ & -$^{\,\,}$\tabularnewline
\object{Ser X--1} & 50507.08 (-0.03) & 34.6$^{\,\,}$ & 45.9$^{\,\,}$\tabularnewline
 & 51399.14 (-0.07) & 309.4$^{\,\,}$ & 1.5$^{\,\,}$\tabularnewline
 & 54753.28 (-0.07) & 54.7$^{\,\,}$ & 0.0$^{\,\,}$\tabularnewline
\hline 
\end{tabular}
\par\end{centering}

$\mathrm{^{a}}$ Time of the observed start of the superburst (see
Sect.~\ref{sec:Introduction} for references). In case this directly
follows a data gap, the duration of the gap is indicated in parenthesis.

$\mathrm{^{b}}$ Using bursts from MINBAR, unless indicated otherwise.

$\mathrm{^{c}}$ Total exposure time in first 35~d after the superburst;
``-'' indicates the superburst occurred outside of the period covered
by the catalog.

$\mathrm{^{d}}$ \citet{Kuulkers2010}.

$\mathrm{^{e}}$ \citet{Keek2008}.

$\mathrm{^{f}}$ \citet{Linares2011ATel}.

$\mathrm{^{g}}$ There were no bursts observed before the subsequent
superburst.

$^{*}$ Superburst took place when hydrogen and helium were likely
burning in steady state.

$\mathrm{^{t}}$ Transient source with outburst durations shorter
than the superburst recurrence time.
\end{table}

Two superbursters are so-called ultracompact X-ray binaries (UCXBs;
e.g., \citealt{intZand2007}), where the accreted material is thought
to be hydrogen deficient, but may contain helium: 4U~0614+091 (\citealt{Kuulkers2010})
and 4U~1820--303 (\citealt{Strohmayer2002,Zand2011ATel}; see also
\citealt{2003Cumming}). From 4U~0614+091 two bursts were observed
in the period after the superburst: after $18.6$ and $34.9$ days,
respectively (\citealt{Kuulkers2010}). Both bursts reach a bolometric
peak flux that is consistent within $1\sigma$. The first burst is
more symmetric with a rise of $5\,\mathrm{s}$ and a decay of $2.0\,\mathrm{s}$,
whereas the second one has a faster rise of $1\,\mathrm{s}$ and a
slower decay of $13.0\,\mathrm{s}$. It should be noted, however,
that the rise and decay time scales were derived from different energy
bands, with the first burst being observed at higher energies, causing
part of the shorter decay time.

\begin{figure*}
\begin{centering}
\includegraphics{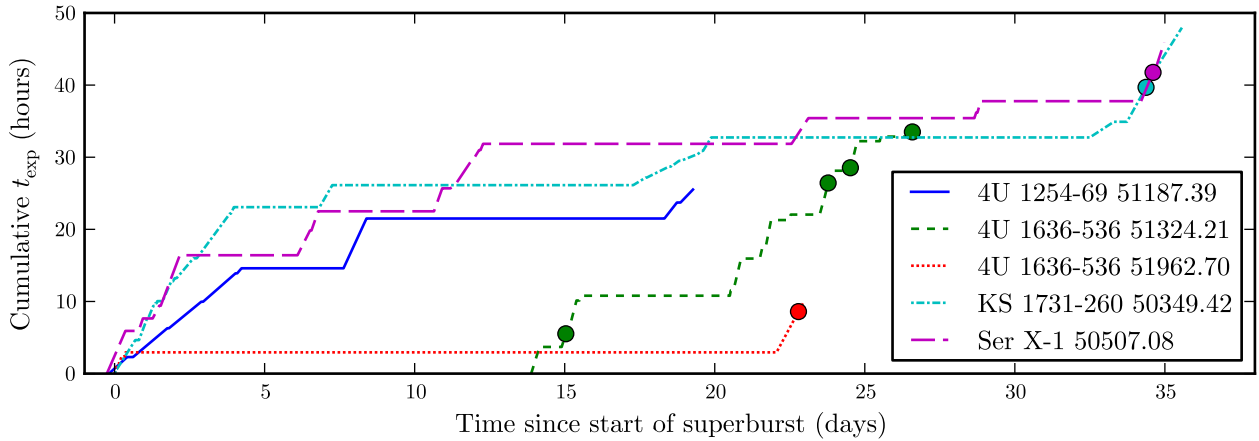}
\par\end{centering}

\caption{\label{fig:exposure}Cumulative net exposure time during 35~days
after a superburst for hydrogen-accreting sources. We exclude sources
with less than $3\,\mathrm{hr}$ of exposure time, and those for which
hydrogen/helium burning is stable, as well as GX~17+2 (see also Table~\ref{tab:Observational-limits-on}).
For each superburst we indicate the source name and the time of the
superburst (MJD). Circles indicate the detection of normal bursts.}
\end{figure*}

The other superbursting sources are thought to accrete hydrogen-rich
material. As the model with the hydrogen-rich envelope predicts normal
bursting behavior to return in $35\,\mathrm{days}$, we identify the
bursts from MINBAR observed in that period, as well as the net exposure
time, $t_{\mathrm{exp}}$, of all observations, including those without
bursts (Fig.~\ref{fig:exposure}, Table~\ref{tab:Observational-limits-on}).
The earliest observed normal burst occurred $15.0$~days after the
superburst from 4U~1636--536 on MJD~$51324.21$. After most superbursts
the total exposure time is short, and observations are infrequent.
Therefore, we combine the data for these sources in $1\,\mathrm{day}$
time bins to be able to distinguish changes in the burst behavior
on similar time scales as exhibited by our models. To place constraints
on the burst recurrence time, $t_{\mathrm{recur}}$, we use the combined
$t_{\mathrm{exp}}$ and number of bursts during each day, taking into
account the presence of data gaps by Earth occultations and by passages
through the South Atlantic Anomaly.

Data gaps reduce the fraction of the observation time $t_{\mathrm{obs}}$
when bursts can be observed: $t_{\mathrm{exp}}=\eta t_{\mathrm{obs}}$,
with $\eta$ the observation efficiency. For $\eta<1$ bursts can
be missed, and when $N$ bursts occur, $n\le N$ are detected. Using
a binomial distribution, the probability $P$ of detecting $n$ bursts
out of $N$ is 
\[
P(n;N,\eta)=\binom{N}{n}\eta^{n}(1-\eta)^{N-n}.
\]
Since $n$ and $\eta$ are known for each observation, we use $P$
to identify those values of $N$ for which the probability of detecting
$n$ is largest. After normalization, we determine the 90\% confidence
region for $N$, which we use to constrain $t_{\mathrm{recur}}=t_{\mathrm{obs}}/N$. 

\begin{figure}
\includegraphics{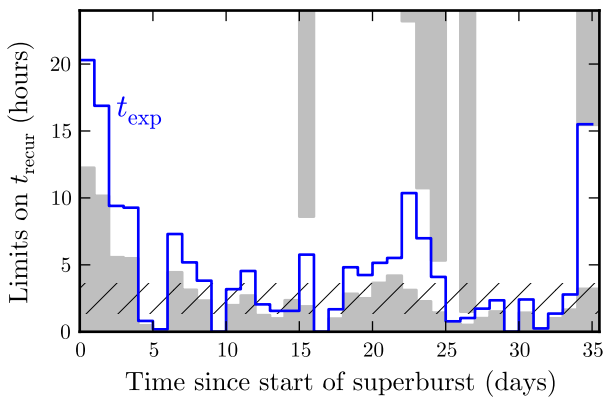}

\caption{\label{fig:trecur_limits}Observational limits on the burst recurrence
time $t_{\mathrm{recur}}$ from combined observations after the superbursts
from Fig.~\ref{fig:exposure} in $1\,\mathrm{day}$ time bins. The
region between the gray areas is the 90\% confidence region for $t_{\mathrm{recur}}$.
The solid line is the net exposure time $t_{\mathrm{exp}}$ per day,
and the hatched region indicates the range of minimum values of $t_{\mathrm{recur}}$
observed for bursts at similar persistent flux as at the time of the
superbursts.}
\end{figure}

As additional constraint, if no bursts are observed, we require that
$t_{\mathrm{recur}}$ is longer than the uninterrupted part of a pointing.
For a $96\,\mathrm{minute}$ satellite orbit we estimate this as $(96\,\mathrm{minutes})\times\eta$.
Furthermore, in our superburst selection only for the superburst on
$51324.21$ from 4U~1636--536 more than one burst was observed in
the following month. The closest pair was separated by $17.8\,\mathrm{hr}$.
We use this as upper limit for $t_{\mathrm{recur}}$ at 24~days after
the superburst onset. The combined constraints are presented in Fig.~\ref{fig:trecur_limits}.

To determine the expected $t_{\mathrm{recur}}$ before the superbursts,
we search MINBAR for the shortest time interval between two subsequent
bursts that occurred at a similar level of persistent flux as the
superburst. For six superbursts we find time intervals between $1.4\,\mathrm{hr}$
and $3.6\,\mathrm{hr}$, whereas for others we find time intervals
of $8.0\,\mathrm{hr}$ and longer. The longer $t_{\mathrm{recur}}$
are most likely because of a short total exposure time on a given
source at the level of persistent flux of interest, and not representative
of the actual $t_{\mathrm{recur}}$. Indeed, the shorter time intervals
are found for most of the superbursts in Fig.~\ref{fig:exposure},
which have the longest cumulative exposure times. Therefore, we take
$1.4$--$3.6\,\mathrm{hr}$ to be the range of expected values for
$t_{\mathrm{recur}}$.

Comparing the expected pre-superburst $t_{\mathrm{recur}}$ to the
limits on $t_{\mathrm{recur}}$ (Fig.~\ref{fig:trecur_limits}),
we see that $t_{\mathrm{recur}}\leq5.5\,\mathrm{hr}$ is strongly
disfavored during the first $4$ days, with $t_{\mathrm{recur}}$
likely exceeding $12.3\,\mathrm{hr}$ during the first day. Therefore,
these superbursts caused a substantial increase in the burst recurrence
time, or possibly quenched bursts altogether. Furthermore, for most
$1\,\mathrm{day}$ time bins the lower limit for $t_{\mathrm{recur}}$
is at least $0.5$--$1.0\,\mathrm{hr}$, which disfavors bursts with
shorter $t_{\mathrm{recur}}$.

The strongest constraints are due to a burst from 4U~1636--536 on
day 26: $0.5\,\mathrm{hr}\leq t_{\mathrm{recur}}\leq1.5\,\mathrm{hr}$.
This suggests $t_{\mathrm{recur}}$ is back at pre-superburst values
after $26$ days. Only $5$ days earlier, the lower limit was $t_{\mathrm{recur}}\ge4.2\,\mathrm{hr}$,
and on day 34 $t_{\mathrm{recur}}\ge3.2\,\mathrm{hr}$: both at the
upper part of the pre-superburst range. Therefore, the allowed values
for $t_{\mathrm{recur}}$ within the pre-superburst range vary somewhat
with time, which may be due to differences in the persistent flux
from one superburst to the next as well as variations during the month
after a given superburst.

The majority of the bursts indicated in Fig.~\ref{fig:exposure}
followed the superburst on MJD~51324.21 from 4U~1636--536. All those
bursts have similar properties, and exhibit PRE, indicating that the
Eddington limit was reached. The persistent flux was approximately
$6\,\%$ of the burst peak flux. The burst properties are typical
for bursts from this source at that level of persistent flux. Burst
properties not changing between days $15$ and $26$ suggest that
normal bursting behavior was resumed at most two weeks after the superburst.

\section{Discussion}

We present one-dimensional multi-zone simulations of the neutron star
envelope, where we study the effect of a superburst on a helium-rich
and on a solar-composition atmosphere, where matter is accreted at
a rate of $\dot{M}=5.25\times10^{-9}\,\mathrm{M_{\odot}}\mathrm{yr^{-1}}$.
The simulations are similar to those presented by \citet{Keek2011}
where a carbon-rich layer is build up by accretion until a superburst
ignites. In the present work we replace the accretion composition
shortly before the superburst to build an atmosphere of either a helium-rich
or a solar composition. The results are summarized in Table~\ref{tab:Superburst-properties-for}.

We find that the hydrogen/helium-rich atmosphere changes the shock
breakout and precursor peak height and duration, whereas the superburst
itself does not differ substantially from models with a carbon-rich
atmosphere. After the superburst all burning in the atmosphere is
stable, and bursts are quenched until the envelope has cooled down
from the superburst. At that point burning first becomes marginally
stable, followed by weak bursts with short recurrence times. Over
time, recurrence times lengthen, and the burst properties return to
those from before the superburst.

\subsection{Superburst}

\citet{Keek2011} found for a model with the same accretion rate and
a slightly higher $Q_{\mathrm{b}}=0.13\,\mathrm{MeV\, nucleon^{-1}}$
a recurrence time of $1.70$~years, $33\,\%$ longer than the recurrence
time found in the present study. With $Q_{\mathrm{b}}=0.1\,\mathrm{MeV\, nucleon^{-1}}$
one would expect our cooler model to produce a somewhat longer recurrence
time. For the models in the current paper, we use a different implementation
of accretion and, most importantly, of the compressional heating from
the accreted material. A further difference is the prescription used
for electron conductivity.

The different phases displayed by the superburst are the same as described
by \citet{Keek2011}, including the shock breakout, the precursor,
and the two-component power-law decay. A difference lies in the height
of the shock breakout peak and the oscillations at the start of the
precursor, where super-Eddington luminosities are reached. The peak
luminosity has a strong dependence on the resolution at the surface
of the model (\citealt{Keek2011}). The bottom of the outer zone of
our models is at a column depth of $y\simeq10^{6}\,\mathrm{g\, cm^{-2}}$,
whereas the photosphere of a neutron star is typically located at
$y\simeq1\,\mathrm{g\, cm^{-2}}$, so it is likely that our simulations
have not resolved this fully. A previous study also finds the shock
breakout to be super-Eddington, but likewise does not fully resolve
the photosphere (\citealt{Weinberg2007}). This underlines the importance
of the treatment of the outer layers during these hydrodynamic events.
Whereas our non-relativistic simulations employ diffusive radiation
transport, a relativistic hydrodynamic model that includes full radiation
transport will be much better suited to simulate these processes in
detail. Such a model will also be able to determine whether any material
is lost from the neutron star during the super-Eddington luminosity
phases.

\subsection{Precursor}

After the initial oscillations, the precursor luminosity settles at
the Eddington limit, $L_{\mathrm{Edd}}$, producing a plateau in the
light curve. The height of the plateau is different for the three
atmosphere compositions because of the dependence of $L_{\mathrm{Edd}}$
on the opacity, $\kappa$: $L_{\mathrm{Edd}}\propto\kappa^{-1}$.
The opacity in the neutron star photosphere is typically assumed to
be dominated by Thomson scattering, which depends exclusively on the
hydrogen mass-fraction. Both models He and C are devoid of hydrogen,
but they reach the Eddington limit at slightly different luminosities.
The opacity in the outer zones of the models continue to increase
toward the surface. It is likely that the true photospheric values
are not yet reached, which may explain the variations in $L_{\mathrm{Edd}}$.

The precursor in model C is predominantly powered by the fallback
of shock-heated material. The fluence of this precursor is about half
that of the precursors in the models He and H, indicating that even
in those cases fallback has a substantial contribution to the precursor,
with hydrogen and helium burning accounting for the rest of the fluence.

The superburst ignited at a relatively early phase in the cycle of
hydrogen/helium bursts. If it had ignited at a later phase, the respective
helium-rich and solar composition columns would have been several
times larger, allowing for somewhat longer precursor bursts. Note
that the brightness of the precursors would not change, as it is set
by the Eddington limit for the atmosphere composition. Precursor bursts
that reach the Eddington limit with durations of several seconds have
been observed in the few instances when the start of the superburst
was observed and the data were of sufficient quality (e.g., \citealt{Strohmayer2002a,Strohmayer2002}).

Previous models of a superburst below a helium atmosphere skipped
the fallback, losing an important energy source for the precursor
(\citealt{Weinberg2007}). The shock breakout left a flat temperature
profile, delaying the burst from helium burning by a thermal time
scale of several seconds. The precursor of model He arrives on a dynamical
time scale of $\sim10^{-5}\,\mathrm{s}$, and is powered by both the
fallback and the subsequent thermonuclear burning of helium. Furthermore,
\citealt{Weinberg2007} argue that the thermonuclear runaway of helium
is triggered either by the shock if the helium layer is sufficiently
thick, or by the carbon deflagration reaching the helium layer. In
our simulations the carbon deflagration does not reach that far out,
and the shock does not trigger the helium burst immediately. It is
rather the fallback that heats the atmosphere sufficiently to burn
helium, even if the helium layer is thin (Fig.~\ref{fig:h_precursor}).

\subsection{Burst Quenching\label{sub:Burst-quenching}}

The superburst heats the atmosphere such that all subsequently accreted
hydrogen and helium burn stably, and no bursts are produced. Burst
quenching continues until the envelope cools down sufficiently for
burning to become unstable again. Using the analytic estimate from
\citet{2004CummingMacBeth}, one predicts for the conditions of our
model a quenching time of $5.5\,\mathrm{days}$ for model H and $2.2\,\mathrm{days}$
for model He. Here it is assumed that $3\alpha$ is solely responsible
for the stability of thermonuclear burning. Model H, however, yields
the shortest quenching time of $1.1\,\mathrm{days}$, whereas in model
He bursts are quenched for $11.3\,\mathrm{days}$. The latter is over
five times longer than the analytic estimate, which may indicate that
the approximations employed in the estimate are too strong. In contrast,
we find a five times \emph{shorter} quenching time for model H. In
this case $3\alpha$ is not the sole process responsible for the resumption
of bursts, but the CNO breakout reaction $^{15}\mathrm{O}(\alpha,\gamma)^{19}\mathrm{Ne}$
plays a key role.

The importance of the $^{15}\mathrm{O}(\alpha,\gamma)^{19}\mathrm{Ne}$
reaction for regulating the CNO abundance in the context of X-ray
bursts was stressed before by \citet{Fisker2006}, and indicated as
an important factor in the stability of the burning processes (\citealt{Fisker2007,Fisker2008}).
Our study shows its importance in determining the duration of the
quenching period. 

Mixing due to rotation, or a rotationally induced magnetic field,
can also influence the stability of thermonuclear burning (\citealt{Piro2007,Keek2009}).
Rotation was not taken into account in this study, but could lead
to a somewhat longer quenching period.

For the ignition column depth of our simulations, $y=1.1\times10^{12}\,\mathrm{g\, cm^{-2}}$,
\citet{Keek2011} showed that the fraction of the superburst energy
that is lost in neutrinos is small, whereas for ignition at larger
depths neutrino losses are substantial, and the photon fluence reaches
a maximum value. This maximum fluence implies that there is a maximum
burst quenching time, which is longer than the values found in our
simulations.

\subsection{Marginally Stable Burning}

When burst quenching ends, burning becomes marginally stable, producing
oscillations in the light curve. This has been observed from a small
number of sources as mHz quasi-periodic oscillations (mHz QPOs; \citealt{Revnivtsev2001,Altamirano2008,Linares2011}),
and is associated with a burning mode at the transition of stable
and unstable burning (\citealt{Heger2005}). Using the one-zone analytic
approximation from Equation (11) in \citet{Heger2005}, and substituting
the values appropriate for model He ($E_{\star}=594\,\mathrm{keV\, nucleon^{-1}}$,
$T_{8}=0.54$, $y_{8}=1.0$, and $\dot{m}=0.3\,\dot{m}_{\mathrm{Edd}}$),
one obtains the oscillation period $P_{\mathrm{osc}}=25\,\mathrm{minutes}$,
which is close to the $20\,\mathrm{minute}$ period in our simulations.
Marginally stable burning in model H has a period of $5\,\mathrm{minutes}$,
which is close to $P_{\mathrm{osc}}=6\,\mathrm{minutes}$ predicted
by the analytic approximation with $T_{8}=4.2$, $y_{8}=0.60$, and
$\dot{m}=0.3\,\dot{m}_{\mathrm{Edd}}$. The analytic estimate from
\citet{Heger2005} considers only accretion, radiative cooling, and
a single thermonuclear process, the rate of which has a similar temperature
dependence as the cooling rate. In the case of model He, where $3\alpha$
is the single dominant thermonuclear process, this approximation seems
valid. It is interesting that it also provides a good estimate of
$P_{\mathrm{osc}}$ in model H, where the nuclear reactions proceed
through a complex interplay between the $3\alpha$, hot-CNO, and CNO
breakout processes.

\subsection{Burst Resumption}

As the envelope continues to cool, the oscillatory burning transitions
into bursts. At first their fluence is relatively low and the recurrences
times are short. Over time (almost four months for model He; one month
for H) the burst properties regain their pre-superburst values. The
light curves of the later bursts, as well as of the pre-superburst
bursts, are qualitatively similar to those reported from recent multi-zone
simulations (e.g., \citealt{Joss1980,Woosley2004,Fisker2008,Jose2010}),
which provide very good agreement with observed bursts (\citealt{Heger2007,Zand2009,Cyburt2010}).

The transition from stable burning, to bursts, via marginally stable
burning (mHz QPOs), and weak bursts, has not been observed directly
after a superburst. In this respect the bursting behavior of \object{IGR~J17480-2446}
in the globular cluster Terzan~5 is interesting (\citealt{Motta2011,Linares2011letter,Chakraborty2011,Linares2011}).
During an outburst in late 2010, its accretion rate varied by more
than a factor five, and it displayed a continuous transition from
bright bursts to weak bursts, to mHz QPOs, and back. The changes in
the burst properties are quite similar to those in model H. The weak
bursts have a longer rise and recurrence times as short as a few minutes.
The bursts become brighter with longer recurrence times and longer
decay profiles. The decrease of the accretion rate causes the atmosphere
to cool, similar to the cooling after the superburst in the models,
but it also changes the burst fuel accumulation time. This is more
complex than the situation in the simulations, and we cannot compare
the time scale on which the changes in burst behavior take place,
as it is set by the variations in the mass accretion rate, instead
of the cooling time scale of the envelope.

\subsection{Carbon Production and Destruction}

One of the biggest challenges in superburst theory is the creation
of the correct amount of carbon. Fits to superburst light curves determine
the carbon mass fraction to be close to $20\,\%$ (\citealt{2004CummingMacBeth,Cumming2006}),
which is the amount we adopted for our models. Models of helium bursts,
or mixed hydrogen/helium bursts, however, produce less than half of
that (e.g., \citealt{Woosley2004}). It is speculated that both bursts
and stable burning of hydrogen and helium are required to produce
enough carbon for superbursts. In fact, most superbursting sources
exhibit this combination of burning behavior (\citealt{Zand2003}). 

Our models exhibit both burning regimes, and the next superburst will
ignite at the bottom of the layer that is accreted just after the
simulated superburst. Because of the relatively high temperature in
the envelope after the superburst, however, most of the produced carbon
burns once it is compressed to higher densities over time. Simulations
over a longer time are required to determine how much carbon can survive
when the envelope cools down further, and the carbon fusion reaction
rate is reduced. Model He might produce a carbon-rich layer where
the next superburst can ignite, whereas in model H most carbon is
destroyed. 

\citet{Medin2011} recently suggested that a larger carbon fraction
can be obtained at the superburst ignition depth by separating out
the carbon from heavier elements by freezing at the crust--ocean interface.

\subsection{Observations after Superbursts}

The detection of a burst is a certain upper limit to the quenching
time. Table~\ref{tab:Observational-limits-on} presents the current
observational upper limits on burst quenching for all known superbursts,
based on the first detection of a burst following each superburst.
We identify an upper limit of $15.0\,\mathrm{days}$ for 4U~1636--536
using the MINBAR burst catalog. This is the shortest reported value
apart from the $2.2\,\mathrm{day}$ limit for GX~17+2, which accretes
at a much higher rate than all other superbursters. 4U~1636--536
accretes hydrogen-rich material at a rate of approximately $0.1\,\dot{M}_{\mathrm{Edd}}$.
The $15.0\,\mathrm{day}$ upper limit is well above the $1.1\,\mathrm{day}$
quenching time we find for model H, which has a three time higher
mass accretion rate. The shortest upper limit for UCXBs, which may
accrete helium-rich material, is $18.6\,\mathrm{days}$ for 4U~0614+091
(\citealt{Kuulkers2010}). Model He has a lower quenching time of
$t_{\mathrm{quench}}=11.3\,\mathrm{days}$, but has a 30 times larger
mass accretion rate.

\citet{2004CummingMacBeth} estimate the dependence of $t_{\mathrm{quench}}$
on the accretion rate to be $t_{\mathrm{quench}}\propto\dot{M}^{-3/4}$.
Using this to correct for the differences in accretion rate between
the models and the observations, model H's $t_{\mathrm{quench}}=6.6\,\mathrm{days}$
is still below the observational upper limit, but model He's $t_{\mathrm{quench}}=28.9\,\mathrm{days}$
lies above the upper limit. Considering, however, that the analytic
approximation does not reproduce the results from our multi-zone models
well, the accretion rate dependence here employed requires further
scrutiny before conclusions can be drawn.

The two bursts following the superburst from 4U~0614+091 have the
same peak flux and suggest a trend towards longer bursts at later
times, just as predicted by model He.

For the hydrogen-accreting sources we combine the observations following
superbursts, that are reported in MINBAR. There is a strong indication
for burst quenching during the first $4$ days after superbursts.
A burst observed after $26\,\mathrm{days}$ firmly constrains the
$t_{\mathrm{recur}}$ at a pre-superburst value. Furthermore, the
lack in variation of the properties of bursts observed from 4U~1636--536
suggests that pre-superburst bursting behavior had returned already
after $15\,\mathrm{days}$, which is shorter than the $35\,\mathrm{days}$
predicted by model H. \citet{Cumming2006} derive from fits to the
light curve of a different superburst from the same source an ignition
column depth that is lower by a factor $0.45$ compared to the ignition
depth in our models. This implies a shorter cooling time scale after
the superburst ($t_{\mathrm{cool}}\propto y_{\mathrm{ign}}^{3/4}$;
\citealt{2004CummingMacBeth}), and hence a faster return to normal
bursts of $19.3\,\mathrm{days}$, which is closer to what we infer
from the observations.

For all but a few days, we derive $t_{\mathrm{recur}}\ge1\,\mathrm{hr}$.
This leaves little room for the short recurrence times predicted by
our models and observed from IGR~J17480-2446. The majority of the
observations during the first week after superbursts were, however,
performed with the \emph{BeppoSAX} WFCs, which did not have enough
sensitivity to detect oscillatory behavior or weak bursts (see, e.g.,
\citealt{Keek2010} for a comparison of the detection of weak bursts
with the \emph{RXTE} PCA and \emph{BeppoSAX} WFCs). Alternatively,
the bursting behavior after a superburst may resemble that close to
the transition to stable burning observed at higher $\dot{M}$, when
the burst rate decreases (e.g., \citealt{Cornelisse2003}).

Both models H and He predict a period when weak and brighter bursts
alternate. This behavior has not been observed, which is not surprising
considering the short time that it is expected to take place. For
model H this burning mode lasts a few hours, whereas for model He
it continues for $2.3$~days. It is, therefore, easiest to observe
this behavior from a UCXB.

Because observations after a superburst are often few and far apart
(Fig.~\ref{fig:exposure}), we have to combine the data for superbursts
which differ by up to a factor $6$ in $y_{\mathrm{ign}}$ (\citealt{Cumming2006}),
and have corresponding different quenching and cooling times. A campaign
of frequent and longer observations during the month following a superburst
would, therefore, be very important in improving our current inability
to accurately predict changes in bursting behavior. Now that \emph{RXTE}
has ceased its operations, there is no observatory available with
a sufficiently large effective area and time resolution to study the
mHz QPOs and weak bursts, as well as the brief superburst precursors.
A future mission such as \emph{LOFT}, that greatly improves on collecting
area, will be able to make important steps forward in our understanding
of thermonuclear processes on neutron stars (\citealt{Feroci2011_LOFT}).

\section{Conclusions}

We create one-dimensional multi-zone models of a neutron star envelope
undergoing a superburst (carbon flash), in the presence of an atmosphere
of either pure helium or of solar composition. The latter is the first
model of its kind. After the superburst we continue the simulations
to study burst quenching as well as the return of normal Type~I X-ray
bursts, simulating over 900 hydrogen or helium flashes per model.

The heating of the atmosphere by fallback of shocked material generates
a precursor burst that reaches the Eddington limit. Any available
hydrogen or helium ignites, which extends the duration of the precursor.

After the superburst, the atmosphere is sufficiently hot for bursts
to be quenched, and all hydrogen and helium to burn in a stable manner.
In a pure helium layer mostly carbon is produced, and in a layer of
solar composition stable \textsl{rp}-process burning creates mostly
germanium and gallium. As the envelope cools down from the superburst,
bursts reappear in the light curve. In the helium atmosphere this
happens after $11.3\,\mathrm{days}$, when the $3\alpha$-process
becomes unstable. We find that in a solar composition quenching ends
much sooner, after only $1.1\,\mathrm{days}$. In this case it is
the interplay between $3\alpha$, the hot CNO cycle, and the $^{15}\mathrm{O}(\alpha,\gamma)^{19}\mathrm{Ne}$
breakout reaction that leads to unstable burning. 

In both models, at the transition from stable burning to unstable
burning, oscillations are produced in the light curve due to marginally
stable burning (\citealt{Heger2005}), followed for a brief time by
alternating weak and brighter bursts, which have longer and shorter
recurrence times, respectively. The latter bursting mode has not been
observed yet. Eventually the weak bursts disappear. In the helium
atmosphere the bright bursts are immediately as bright as those before
the superburst, whereas in the solar atmosphere burst peak luminosities
grow as the envelope cools further. For a few days left-over helium
above the burst ignition column depth burns, pausing the cooling.
Afterward the burst durations and recurrence times increase back to
the pre-superburst values over the course of one month for solar composition,
and four months for helium composition.

The transition from burst quenching to bursts after a superburst has
not been directly observed. Using the MINBAR catalog we identify the
shortest reported upper limit to the quenching time of $15.0\,\mathrm{days}$
for 4U~1636--536 (with the exception of GX~17+2), and we derive
further constraints on the time scales for quenching and the return
of bursts. The short recurrence times found by the simulations are
disfavored, but not excluded. The transition between the different
burning regimes that we describe exhibits, however, strong similarities
with bursts observed from the transient burster IGR~J17480--2446
in Terzan~5 (e.g., \citealt{Linares2011}).

\acknowledgements{The authors thank K. Chen for helpful discussions. This paper uses
preliminary analysis results from the Multi-INstrument Burst ARchive
(MINBAR), which is supported under the Australian Academy of Science's
Scientific Visits to Europe program, and the Australian Research Council's
Discovery Projects and Future Fellowship funding schemes. The authors
thank the International Space Science Institute in Bern for hosting
an International Team on Type I X-ray bursts. L.K. is supported by
the Joint Institute for Nuclear Astrophysics (JINA, grant PHY08-22648),
a National Science Foundation Physics Frontier Center. A.H. acknowledges
support from the DOE Program for Scientific Discovery through Advanced
Computing (SciDAC, DE-FC02-09ER41618) and by the US Department of
Energy under grant DE-FG02-87ER40328.}

\bibliographystyle{apj}
\bibliography{apj-jour,sb_hhe}

\begin{thebibliography}{62}
\expandafter\ifx\csname natexlab\endcsname\relax\def\natexlab#1{#1}\fi

\bibitem[{{Altamirano} {et~al.}(2008){Altamirano}, {van der Klis}, {Wijnands},
  \& {Cumming}}]{Altamirano2008}
{Altamirano}, D., {van der Klis}, M., {Wijnands}, R., \& {Cumming}, A. 2008,
  \apjl, 673, L35

\bibitem[{{Asada} {et~al.}(2011){Asada}, {Negoro}, {Sugizaki}, {Matsuoka},
  {Mihara}, {Serino}, {Nakahira}, {Yamamoto}, {Sootome}, {Ueno}, {Tomida},
  {Kohama}, {Ishikawa}, {Kawai}, {Morii}, {Sugimori}, {Usui}, {Toizumi},
  {Aoki}, {Song}, {Yoshida}, {Yamaoka}, {Tsunemi}, {Kimura}, {Kitayama},
  {Nakajima}, {Suwa}, {Sakakibara}, {Ueda}, {Hiroi}, {Shidatsu}, {Tsuboi},
  {Matsumura}, {Yamauchi}, {Nishimura}, \& {Hanayama}}]{Asada2011ATel}
{Asada}, M., {et~al.} 2011, The Astronomer's Telegram, 3760, 1

\bibitem[{{Caughlan} \& {Fowler}(1988)}]{Caughlan1988}
{Caughlan}, G.~R., \& {Fowler}, W.~A. 1988, Atomic Data and Nuclear Data
  Tables, 40, 283

\bibitem[{{Chakraborty} \& {Bhattacharyya}(2011)}]{Chakraborty2011}
{Chakraborty}, M., \& {Bhattacharyya}, S. 2011, \apjl, 730, L23

\bibitem[{{Chenevez} {et~al.}(2011){Chenevez}, {Brandt}, {Kuulkers},
  {Alfonso-Garzon}, {Beckmann}, {Bird}, {Courvoisier}, {Del Santo}, {Domingo},
  {Ebisawa}, {Jonker}, {Kretschmar}, {Markwardt}, {Oosterbroek}, {Paizis},
  {Pottschmidt}, {Sanchez-Fernandez}, \& {Wijnands}}]{Chenevez2011ATel}
{Chenevez}, J., {et~al.} 2011, The Astronomer's Telegram, 3183, 1

\bibitem[{{Clayton}(1968)}]{Clayton1968book}
{Clayton}, D.~D. 1968, {Principles of stellar evolution and nucleosynthesis},
  ed. {Clayton, D.~D.} (New York: McGraw-Hill, 1968)

\bibitem[{{Cooper} {et~al.}(2009){Cooper}, {Steiner}, \& {Brown}}]{Cooper2009}
{Cooper}, R.~L., {Steiner}, A.~W., \& {Brown}, E.~F. 2009, \apj, 702, 660

\bibitem[{{Cornelisse} {et~al.}(2000){Cornelisse}, {Heise}, {Kuulkers},
  {Verbunt}, \& {in~'t~Zand}}]{Cornelisse2000}
{Cornelisse}, R., {Heise}, J., {Kuulkers}, E., {Verbunt}, F., \& {in~'t~Zand},
  J.~J.~M. 2000, \aap, 357, L21

\bibitem[{{Cornelisse} {et~al.}(2003){Cornelisse}, {in~'t~Zand}, {Verbunt},
  {Kuulkers}, {Heise}, {den Hartog}, {Cocchi}, {Natalucci}, {Bazzano}, \&
  {Ubertini}}]{Cornelisse2003}
{Cornelisse}, R., {et~al.} 2003, \aap, 405, 1033

\bibitem[{{Cumming}(2003)}]{2003Cumming}
{Cumming}, A. 2003, \apj, 595, 1077

\bibitem[{{Cumming} \& {Bildsten}(2001)}]{Cumming2001}
{Cumming}, A., \& {Bildsten}, L. 2001, \apjl, 559, L127

\bibitem[{{Cumming} \& {Macbeth}(2004)}]{2004CummingMacBeth}
{Cumming}, A., \& {Macbeth}, J. 2004, \apjl, 603, L37

\bibitem[{{Cumming} {et~al.}(2006){Cumming}, {Macbeth}, {in~'t~Zand}, \&
  {Page}}]{Cumming2006}
{Cumming}, A., {Macbeth}, J., {in~'t~Zand}, J.~J.~M., \& {Page}, D. 2006, \apj,
  646, 429

\bibitem[{{Cyburt} {et~al.}(2010){Cyburt}, {Amthor}, {Ferguson}, {Meisel},
  {Smith}, {Warren}, {Heger}, {Hoffman}, {Rauscher}, {Sakharuk}, {Schatz},
  {Thielemann}, \& {Wiescher}}]{Cyburt2010}
{Cyburt}, R.~H., {et~al.} 2010, \apjs, 189, 240

\bibitem[{{Feroci} {et~al.}(2011){Feroci}, {Stella}, {van der Klis},
  {Courvoisier}, {Hernanz}, {Hudec}, {Santangelo}, {Walton}, {Zdziarski},
  {Barret}, {Belloni}, {Braga}, {Brandt}, {Budtz-J{\o}rgensen}, {Campana}, {den
  Herder}, {Huovelin}, {Israel}, {Pohl}, {Ray}, {Vacchi}, {Zane}, {Argan},
  {Attin{\`a}}, {Bertuccio}, {Bozzo}, {Campana}, {Chakrabarty}, {Costa}, {de
  Rosa}, {Del Monte}, {di Cosimo}, {Donnarumma}, {Evangelista}, {Haas},
  {Jonker}, {Korpela}, {Labanti}, {Malcovati}, {Mignani}, {Muleri},
  {Rapisarda}, {Rashevsky}, {Rea}, {Rubini}, {Tenzer}, {Wilson-Hodge},
  {Winter}, {Wood}, {Zampa}, {Zampa}, {Abramowicz}, {Alpar}, {Altamirano},
  {Alvarez}, {Amati}, {Amoros}, {Antonelli}, {Artigue}, {Azzarello},
  {Bachetti}, {Baldazzi}, {Barbera}, {Barbieri}, {Basa}, {Baykal}, {Belmont},
  {Boirin}, {Bonvicini}, {Burderi}, {Bursa}, {Cabanac}, {Cackett}, {Caliandro},
  {Casella}, {Chaty}, {Chenevez}, {Coe}, {Collura}, {Corongiu}, {Covino},
  {Cusumano}, {D'Amico}, {Dall'Osso}, {de Martino}, {de Paris}, {di Persio},
  {di Salvo}, {Done}, {Dov{\v c}iak}, {Drago}, {Ertan}, {Fabiani}, {Falanga},
  {Fender}, {Ferrando}, {Della Monica Ferreira}, {Fraser}, {Frontera},
  {Fuschino}, {Galvez}, {Gandhi}, {Giommi}, {Godet}, {G{\"o}{\v g}{\"u}{\c s}},
  {Goldwurm}, {G{\"o}tz}, {Grassi}, {Guttridge}, {Hakala}, {Henri}, {Hermsen},
  {Horak}, {Hornstrup}, {in't Zand}, {Isern}, {Kalemci}, {Kanbach}, {Karas},
  {Kataria}, {Kennedy}, {Klochkov}, {Klu{\'z}niak}, {Kokkotas}, {Kreykenbohm},
  {Krolik}, {Kuiper}, {Kuvvetli}, {Kylafis}, {Lattimer}, {Lazzarotto}, {Leahy},
  {Lebrun}, {Lin}, {Lund}, {Maccarone}, {Malzac}, {Marisaldi}, {Martindale},
  {Mastropietro}, {McClintock}, {McHardy}, {Mendez}, {Mereghetti}, {Miller},
  {Mineo}, {Morelli}, {Morsink}, {Motch}, {Motta}, {Mu{\~n}oz-Darias},
  {Naletto}, {Neustroev}, {Nevalainen}, {Olive}, {Orio}, {Orlandini},
  {Orleanski}, {Ozel}, {Pacciani}, {Paltani}, {Papadakis}, {Papitto},
  {Patruno}, {Pellizzoni}, {Petr{\'a}{\v c}ek}, {Petri}, {Petrucci}, {Phlips},
  {Picolli}, {Possenti}, {Psaltis}, {Rambaud}, {Reig}, {Remillard},
  {Rodriguez}, {Romano}, {Romanova}, {Schanz}, {Schmid}, {Segreto}, {Shearer},
  {Smith}, {Smith}, {Soffitta}, {Stergioulas}, {Stolarski}, {Stuchlik},
  {Tiengo}, {Torres}, {T{\"o}r{\"o}k}, {Turolla}, {Uttley}, {Vaughan},
  {Vercellone}, {Waters}, {Watts}, {Wawrzaszek}, {Webb}, {Wilms}, {Zampieri},
  {Zezas}, \& {Ziolkowski}}]{Feroci2011_LOFT}
{Feroci}, M., {et~al.} 2011, Experimental Astronomy, 1

\bibitem[{{Fisker} {et~al.}(2006){Fisker}, {G{\"o}rres}, {Wiescher}, \&
  {Davids}}]{Fisker2006}
{Fisker}, J.~L., {G{\"o}rres}, J., {Wiescher}, M., \& {Davids}, B. 2006, \apj,
  650, 332

\bibitem[{{Fisker} {et~al.}(2008){Fisker}, {Schatz}, \&
  {Thielemann}}]{Fisker2008}
{Fisker}, J.~L., {Schatz}, H., \& {Thielemann}, F.-K. 2008, \apjs, 174, 261

\bibitem[{{Fisker} {et~al.}(2007){Fisker}, {Tan}, {G{\"o}rres}, {Wiescher}, \&
  {Cooper}}]{Fisker2007}
{Fisker}, J.~L., {Tan}, W., {G{\"o}rres}, J., {Wiescher}, M., \& {Cooper},
  R.~L. 2007, \apj, 665, 637

\bibitem[{{Galloway} {et~al.}(2008){Galloway}, {Muno}, {Hartman}, {Psaltis}, \&
  {Chakrabarty}}]{Galloway2008catalog}
{Galloway}, D.~K., {Muno}, M.~P., {Hartman}, J.~M., {Psaltis}, D., \&
  {Chakrabarty}, D. 2008, \apjs, 179, 360

\bibitem[{{Gupta} {et~al.}(2007){Gupta}, {Brown}, {Schatz}, {M{\"o}ller}, \&
  {Kratz}}]{Gupta2007}
{Gupta}, S., {Brown}, E.~F., {Schatz}, H., {M{\"o}ller}, P., \& {Kratz}, K.-L.
  2007, \apj, 662, 1188

\bibitem[{{Haensel} \& {Zdunik}(1990)}]{Haensel1990}
{Haensel}, P., \& {Zdunik}, J.~L. 1990, \aap, 227, 431

\bibitem[{{Heger} {et~al.}(2007{\natexlab{a}}){Heger}, {Cumming}, {Galloway},
  \& {Woosley}}]{Heger2007}
{Heger}, A., {Cumming}, A., {Galloway}, D.~K., \& {Woosley}, S.~E.
  2007{\natexlab{a}}, \apjl, 671, L141

\bibitem[{{Heger} {et~al.}(2007{\natexlab{b}}){Heger}, {Cumming}, \&
  {Woosley}}]{Heger2005}
{Heger}, A., {Cumming}, A., \& {Woosley}, S.~E. 2007{\natexlab{b}}, \apj, 665,
  1311

\bibitem[{{Iben}(1975)}]{Iben1975}
{Iben}, Jr., I. 1975, \apj, 196, 525

\bibitem[{{in 't Zand} {et~al.}(2011){in 't Zand}, {Serino}, {Kawai}, \&
  {Heinke}}]{Zand2011ATel}
{in 't Zand}, J., {Serino}, M., {Kawai}, N., \& {Heinke}, C. 2011, The
  Astronomer's Telegram, 3625, 1

\bibitem[{{in~'t~Zand} {et~al.}(2004){in~'t~Zand}, {Cornelisse}, \&
  {Cumming}}]{Zand2004}
{in~'t~Zand}, J.~J.~M., {Cornelisse}, R., \& {Cumming}, A. 2004, \aap, 426, 257

\bibitem[{{in~'t~Zand} {et~al.}(2007){in~'t~Zand}, {Jonker}, \&
  {Markwardt}}]{intZand2007}
{in~'t~Zand}, J.~J.~M., {Jonker}, P.~G., \& {Markwardt}, C.~B. 2007, \aap, 465,
  953

\bibitem[{{in~'t~Zand} {et~al.}(2009){in~'t~Zand}, {Keek}, {Cumming}, {Heger},
  {Homan}, \& {M{\'e}ndez}}]{Zand2009}
{in~'t~Zand}, J.~J.~M., {Keek}, L., {Cumming}, A., {Heger}, A., {Homan}, J., \&
  {M{\'e}ndez}, M. 2009, \aap, 497, 469

\bibitem[{{in~'t~Zand} {et~al.}(2003){in~'t~Zand}, {Kuulkers}, {Verbunt},
  {Heise}, \& {Cornelisse}}]{Zand2003}
{in~'t~Zand}, J.~J.~M., {Kuulkers}, E., {Verbunt}, F., {Heise}, J., \&
  {Cornelisse}, R. 2003, \aap, 411, L487

\bibitem[{{Itoh} {et~al.}(1996){Itoh}, {Hayashi}, {Nishikawa}, \&
  {Kohyama}}]{Itoh1996}
{Itoh}, N., {Hayashi}, H., {Nishikawa}, A., \& {Kohyama}, Y. 1996, \apjs, 102,
  411

\bibitem[{{Jos{\'e}} {et~al.}(2010){Jos{\'e}}, {Moreno}, {Parikh}, \&
  {Iliadis}}]{Jose2010}
{Jos{\'e}}, J., {Moreno}, F., {Parikh}, A., \& {Iliadis}, C. 2010, \apjs, 189,
  204

\bibitem[{{Joss} \& {Li}(1980)}]{Joss1980}
{Joss}, P.~C., \& {Li}, F.~K. 1980, \apj, 238, 287

\bibitem[{{Keek} {et~al.}(2010){Keek}, {Galloway}, {in 't Zand}, \&
  {Heger}}]{Keek2010}
{Keek}, L., {Galloway}, D.~K., {in 't Zand}, J.~J.~M., \& {Heger}, A. 2010,
  \apj, 718, 292

\bibitem[{{Keek} \& {Heger}(2011)}]{Keek2011}
{Keek}, L., \& {Heger}, A. 2011, \apj, 743, 189

\bibitem[{{Keek} \& {in~'t~Zand}(2008)}]{Keek2008int..work}
{Keek}, L., \& {in~'t~Zand}, J.~J.~M. 2008, in Proceedings of the 7th INTEGRAL
  Workshop. 8 - 11 September 2008 Copenhagen, Denmark. Online at
  http://pos.sissa.it/cgi-bin/reader/conf.cgi?confid=67, p.32

\bibitem[{{Keek} {et~al.}(2008){Keek}, {in~'t~Zand}, {Kuulkers}, {Cumming},
  {Brown}, \& {Suzuki}}]{Keek2008}
{Keek}, L., {in~'t~Zand}, J.~J.~M., {Kuulkers}, E., {Cumming}, A., {Brown},
  E.~F., \& {Suzuki}, M. 2008, \aap, 479, 177

\bibitem[{{Keek} {et~al.}(2009){Keek}, {Langer}, \& {in~'t~Zand}}]{Keek2009}
{Keek}, L., {Langer}, N., \& {in~'t~Zand}, J.~J.~M. 2009, \aap, 502, 871

\bibitem[{Kuulkers(2004)}]{Kuulkers2003a}
Kuulkers, E. 2004, Nucl. Phys. Proc. Suppl., 132, 466

\bibitem[{{Kuulkers}(2009)}]{Kuulkers2009ATel}
{Kuulkers}, E. 2009, The Astronomer's Telegram, 2140, 1

\bibitem[{{Kuulkers} {et~al.}(2010){Kuulkers}, {in 't Zand}, {Atteia},
  {Levine}, {Brandt}, {Smith}, {Linares}, {Falanga},
  {S{\'a}nchez-Fern{\'a}ndez}, {Markwardt}, {Strohmayer}, {Cumming}, \&
  {Suzuki}}]{Kuulkers2010}
{Kuulkers}, E., {et~al.} 2010, \aap, 514, A65+

\bibitem[{{Kuulkers} {et~al.}(2002){Kuulkers}, {in~'t~Zand}, {van Kerkwijk},
  {Cornelisse}, {Smith}, {Heise}, {Bazzano}, {Cocchi}, {Natalucci}, \&
  {Ubertini}}]{Kuulkers2002ks1731}
---. 2002, \aap, 382, 503

\bibitem[{{Linares} {et~al.}(2012){Linares}, {Altamirano}, {Chakrabarty},
  {Cumming}, \& {Keek}}]{Linares2011}
{Linares}, M., {Altamirano}, D., {Chakrabarty}, D., {Cumming}, A., \& {Keek},
  L. 2012, \apj, 748, 82

\bibitem[{{Linares} {et~al.}(2011{\natexlab{a}}){Linares}, {Altamirano},
  {Watts}, {van der Klis}, {Wijnands}, {Patruno}, {Armas-Padilla}, {Cavecchi},
  {Degenaar}, {Kalamkar}, {Kaur}, {Yang}, {Casella}, \&
  {Rea}}]{Linares2011ATel}
{Linares}, M., {et~al.} 2011{\natexlab{a}}, The Astronomer's Telegram, 3217, 1

\bibitem[{{Linares} {et~al.}(2011{\natexlab{b}}){Linares}, {Chakrabarty}, \&
  {van der Klis}}]{Linares2011letter}
{Linares}, M., {Chakrabarty}, D., \& {van der Klis}, M. 2011{\natexlab{b}},
  \apjl, 733, L17

\bibitem[{{Medin} \& {Cumming}(2011)}]{Medin2011}
{Medin}, Z., \& {Cumming}, A. 2011, \apj, 730, 97

\bibitem[{{Motta} {et~al.}(2011){Motta}, {D'A{\i}}, {Papitto}, {Riggio}, {di
  Salvo}, {Burderi}, {Belloni}, {Stella}, \& {Iaria}}]{Motta2011}
{Motta}, S., {et~al.} 2011, \mnras, 414, 1508

\bibitem[{{Piro} \& {Bildsten}(2007)}]{Piro2007}
{Piro}, A.~L., \& {Bildsten}, L. 2007, \apj, 663, 1252

\bibitem[{{Rauscher} {et~al.}(2003){Rauscher}, {Heger}, {Hoffman}, \&
  {Woosley}}]{Rauscher2003}
{Rauscher}, T., {Heger}, A., {Hoffman}, R.~D., \& {Woosley}, S.~E. 2003,
  Nuclear Physics A, 718, 463

\bibitem[{{Revnivtsev} {et~al.}(2001){Revnivtsev}, {Churazov}, {Gilfanov}, \&
  {Sunyaev}}]{Revnivtsev2001}
{Revnivtsev}, M., {Churazov}, E., {Gilfanov}, M., \& {Sunyaev}, R. 2001, \aap,
  372, 138

\bibitem[{{Schatz} {et~al.}(2001){Schatz}, {Aprahamian}, {Barnard}, {Bildsten},
  {Cumming}, {Ouellette}, {Rauscher}, {Thielemann}, \& {Wiescher}}]{Schatz2001}
{Schatz}, H., {et~al.} 2001, Physical Review Letters, 86, 3471

\bibitem[{{Schatz} {et~al.}(2003){Schatz}, {Bildsten}, {Cumming}, \&
  {Ouelette}}]{2003Schatz}
{Schatz}, H., {Bildsten}, L., {Cumming}, A., \& {Ouelette}, M. 2003, Nuclear
  Physics A, 718, 247

\bibitem[{{Serino} {et~al.}(2012){Serino}, {Mihara}, {Matsuoka}, {Nakahira},
  {Sugizaki}, {Ueda}, {Kawai}, \& {Ueno}}]{Serino2012}
{Serino}, M., {Mihara}, T., {Matsuoka}, M., {Nakahira}, S., {Sugizaki}, M.,
  {Ueda}, Y., {Kawai}, N., \& {Ueno}, S. 2012, ArXiv e-prints

\bibitem[{{Strohmayer} \& {Brown}(2002)}]{Strohmayer2002}
{Strohmayer}, T.~E., \& {Brown}, E.~F. 2002, \apj, 566, 1045

\bibitem[{{Strohmayer} \& {Markwardt}(2002)}]{Strohmayer2002a}
{Strohmayer}, T.~E., \& {Markwardt}, C.~B. 2002, \apj, 577, 337

\bibitem[{{Taam} {et~al.}(1996){Taam}, {Woosley}, \& {Lamb}}]{Taam1996}
{Taam}, R.~E., {Woosley}, S.~E., \& {Lamb}, D.~Q. 1996, \apj, 459, 271

\bibitem[{{Wallace} \& {Woosley}(1981)}]{Wallace1981}
{Wallace}, R.~K., \& {Woosley}, S.~E. 1981, \apjs, 45, 389

\bibitem[{{Weaver} {et~al.}(1978){Weaver}, {Zimmerman}, \&
  {Woosley}}]{Weaver1978}
{Weaver}, T.~A., {Zimmerman}, G.~B., \& {Woosley}, S.~E. 1978, \apj, 225, 1021

\bibitem[{{Weinberg} \& {Bildsten}(2007)}]{Weinberg2007}
{Weinberg}, N.~N., \& {Bildsten}, L. 2007, \apj, 670, 1291

\bibitem[{{Weinberg} {et~al.}(2006){Weinberg}, {Bildsten}, \&
  {Brown}}]{Weinberg2006sb}
{Weinberg}, N.~N., {Bildsten}, L., \& {Brown}, E.~F. 2006, \apjl, 650, L119

\bibitem[{{Woosley} {et~al.}(2004){Woosley}, {Heger}, {Cumming}, {Hoffman},
  {Pruet}, {Rauscher}, {Fisker}, {Schatz}, {Brown}, \&
  {Wiescher}}]{Woosley2004}
{Woosley}, S.~E., {et~al.} 2004, \apjs, 151, 75

\bibitem[{{Woosley} {et~al.}(2002){Woosley}, {Heger}, \&
  {Weaver}}]{Woosley2002RvMP}
{Woosley}, S.~E., {Heger}, A., \& {Weaver}, T.~A. 2002, Reviews of Modern
  Physics, 74, 1015

\bibitem[{{Woosley} \& {Weaver}(1984)}]{Woosley1984}
{Woosley}, S.~E., \& {Weaver}, T.~A. 1984, in American Institute of Physics
  Conference Series, Vol. 115, AIP Conf Ser 115: High Energy Transients in
  Astrophysics, ed. {S.~E.~Woosley}, 273--297

\end{thebibliography}

\end{document}